\documentclass[11pt]{article}
\usepackage{comment}
\usepackage[margin=2.5cm]{geometry}
\usepackage{amsmath,amssymb,extarrows,mathtools,graphicx,subfigure,setspace,bbold}
\usepackage{cite}
\usepackage{slashed}
\usepackage[dvipsnames]{xcolor}
\usepackage{hyperref}
\hypersetup{colorlinks=true, linkcolor=blue, citecolor=magenta, linktoc=page}
\usepackage{tikz}
\usepackage{arydshln,leftidx,mathtools}
\usetikzlibrary{decorations.pathreplacing}
\numberwithin{equation}{section}
\newcommand{\be}{\begin{equation}}
\newcommand{\bea}{\begin{eqnarray}}
\newcommand{\eea}{\end{eqnarray}}
\newcommand{\ba}{\begin{align}}
\newcommand{\ea}{\end{align}}
\newcommand{\ee}{\end{equation}}

\definecolor{myorange}{RGB}{255, 114,0}
\definecolor{myblue}{RGB}{1,1,255}

\begin{document}

\begin{titlepage}
\thispagestyle{empty}

\begin{flushright}
IPM/P-2017/071\\
\end{flushright}

\vspace{.4cm}
\begin{center}
\noindent{\Large \textbf{Logarithmic Negativity in Lifshitz Harmonic Models}}\\

\vspace{2cm}

M. Reza Mohammadi Mozaffar and Ali Mollabashi
\vspace{1cm}

{\it School of Physics}
\\
{\it Institute for Research in Fundamental Sciences (IPM), Tehran, Iran}\\
\vspace{1cm}
Emails: {\tt m$_{-}$mohammadi, mollabashi@ipm.ir}

\vskip 2em
\end{center}

\vspace{.5cm}
\begin{abstract}
Recently generalizations of the harmonic lattice model has been introduced as a discrete approximation of bosonic field theories with Lifshitz symmetry with a generic dynamical exponent $z$. In such models in $(1+1)$ and $(2+1)$-dimensions, we study logarithmic negativity in the vacuum state and also finite temperature states. We investigate various features of logarithmic negativity such as the universal term, its $z$-dependence and also its temperature dependence in various configurations. We present both analytical and numerical evidences for linear $z$-dependence of logarithmic negativity in almost all range of parameters both in $(1+1)$ and $(2+1)$-dimensions. We also investigate the validity of area law behavior of logarithmic negativity in these generalized models and find that this behavior is still correct for small enough dynamical exponents. 
\end{abstract}
\end{titlepage}

\newpage
\tableofcontents
\noindent
\hrulefill
\onehalfspacing

\section{Introduction}
Entanglement structure of many-body states has gained lots of attention in recent years. Specifically in generic space-time dimensions the ground state entanglement entropy captures universal information about intrinsic properties of the theory (through its universal terms) or even information about the universality class of critical models through its scaling properties\cite{Calabrese:2009qy,Laflorencie:2015eck}.

So far the entanglement structure of systems in a pure state considering a bipartite decomposition of the Hilbert space has been widely studied. More precisely for a system in a pure state if we consider a Hilbert space decomposition as $\mathcal{H}=\mathcal{H}_A\otimes\mathcal{H}_{\bar{A}}$ where $\bar{A}$ is the complement of $A$, the entanglement entropy in this case is defined as $S_A=-\mathrm{Tr}\left[\rho_A\log\rho_A\right]$. Another family of important quantities which has been widely studied in such systems are Renyi entropies defined as
$$S^{(n)}_A=\frac{1}{1-n}\log\mathrm{Tr}\rho^n_A.$$
In the context of field theories, entanglement (Renyi) entropies are divergent quantities and we have to introduce a UV cutoff which we will denote by $\epsilon$, in order to regularize these entropies. While we consider $\epsilon\to0$, it is well known that for local theories the most divergent term of entanglement (Renyi) entropies scales with the area of the boundary of region $A$\cite{Srednicki:1993im}. There are some exceptions for this general behavior, the most important ones are critical systems in (1+1)-dimensions, where the entropies scale logarithmically\cite{Holzhey:1994we,Calabrese:2004eu}. The divergent structure in higher dimensions is more rich and there may be universal information encoded in a specific family of divergent term known as universal terms\cite{Casini:2009sr}. 

In this context universal terms mean those which do not depend on the regularization scheme. These terms are mostly studied in the context of conformal field theories both in field theory and also using the Ryu-Takayanagi formula in the context of AdS/CFT correspondence (see e.g. \cite{Rangamani:2016dms} for review). The universal terms contain specific informations about coefficients of trace anomaly in such theories and in general the details depend on the geometry of the entangling region. For example in (2+1)-dimensions universal (logarithmic) terms may be present in the entropy expansion if there are singular points in the boundary of the entangling region\cite{Casini:2006hu}. 

While the system under study is in a pure state, entanglement (Renyi) entropies are a suitable measure to quantify the amount of entanglement between complementary subsystems. In the case when the system is described with a mixed state, entanglement (Renyi) entropies (and all linear combinations of them) fail to be suitable measures for studying the entanglement structure of the system since they mix up classical and quantum correlations. A well known alternative measure to study entanglement in such cases is logarithmic negativity introduced in \cite{Vidal}. Logarithmic negativity is not the only measure introduced for quantifying quantum entanglement in mixed states but is the only well known measure which is computable for many-body systems with almost all expected features except that it does not reduce to von-Neumann entropy for pure states in general.\footnote{In particular in \cite{Yichen Huang} it has been proved that computing a large class of entanglement measures is NP-hard while the logarithmic negativity is an exception.} Also note that the logarithmic negativity is a proper entanglement monotone decreasing under LOCC which is not convex\cite{Plenio:2005cwa}.

In order to define entanglement negativity in a many-body system, lets take $A$ to be an extended system. The system $A$ may be the whole spatial manifold of the theory in a mixed state or a part of the spatial manifold in a pure state (which its complement $\bar{A}$ is traced out). In the latter case we divide $A$ into two parts $A_1$ and $A_2$, i.e., $\mathcal{H}_A=\mathcal{H}_1\otimes\mathcal{H}_2$ where $\mathcal{H}_{1,2}$ are the corresponding Hilbert spaces for $A_{1,2}$. Lets consider $\{|\phi_i^{(1)}\rangle\}$ and $\{|\phi_i^{(2)}\rangle\}$ as a basis for $\mathcal{H}_1$ and $\mathcal{H}_2$ respectively. Now define partial transpose of the density matrix describing $A$ with respect to $A_2$ as
\be
\langle\phi_i^{(1)},\phi_j^{(2)}|\rho_A^{T_2}|\phi_m^{(1)},\phi_n^{(2)}\rangle
\equiv
\langle\phi_i^{(1)},\phi_n^{(2)}|\rho_A|\phi_m^{(1)},\phi_j^{(2)}\rangle.
\ee
The key point is that since the spectrum of $\rho_A^{T_2}$ contains both positive and negative eigenvalues, the logarithmic negativity is defined using the trace norm of this operator as
\be
\mathcal{E}=\log \left|\left|\rho_A^{T_2}\right|\right|.
\ee
This quantity is clearly independent of the basis chosen for the $\mathcal{H}_i$'s and is computable in the context of many-body systems. As we mentioned before, there are several ways which a (sub)system may be described by a mixed state. Here we are interested in two specific constructions: 1) The system may be initially described by a mixed state, e.g., a thermal state, or 2) it can be subsystem of a larger one which is in a pure state.

Logarithmic negativity has been previously studied in extended systems in (1+1)-dimension. This has been done for both bosonic systems and also in fermionic models for several configurations in \cite{Calabrese:2012ew, Calabrese:2012nk, DeNobili:2015dla, Simon, Audenaert, Eisler2, Coser:2015mta, Coser:2015eba, Ruggiero:2016aqr, Herzog:2016ohd, wen, Hoogeveen:2014bqa}.\footnote{For a specific example in a spin-1 model with dynamical exponent see \cite{Chen:2017txi}.} There are also some related studies in (2+1)-dimensional many-body systems \cite{viktor, DeNobili:2016nmj, Kim:2016bzu, Wen:2016bla}.

The goal of this paper is studying this entanglement measure in a generalization of harmonic lattice models known as Lifshitz harmonic models \cite{MohammadiMozaffar:2017nri, He:2017wla} in (1+1) and (2+1)-dimensions\footnote{Entanglement entropy for a free scalar theory with Lifshitz scaling has been recently studied in \cite{Gentle:2017ywk}, where the authors have used holographic entanglement entropy proposal in a proposed Lifshitz geometry originated from free scalar cMERA with Lifshitz symmetry.}. We introduce these family of models in the following of this section. Most of our study is devoted to a nearly massless regime where the model is supposed to be a discretized version of a scalar field theory with Lifshitz scaling symmetry. We are interested in both the vacuum state and also thermal states in these models.

The organization of this paper is as follows: in the following of this section we give a very short review of Lifshitz harmonic lattice and also the prescription how to compute partial transpose in the context of Gaussian models. Section \ref{sec:onepone} is dedicated to results in $(1+1)$-dimensions. In this section we present numerical study of logarithmic negativity in the vacuum state for generic intervals and also analytical results for a specific configuration called $p$-alternating sublattice. We also study logarithmic negativity in thermal states. In section \ref{sec:twopone} we will generalize our study to $(2+1)$-dimensions and specifically study the expected area law for logarithmic negativity and its dependence on the dynamical exponent. The last section is devoted to concluding remarks and some related discussions.

\subsection{Lifshitz Harmonic Lattice}
The ``harmonic lattice" is well known to be the discretized version of free scalar field theory with Lorentz symmetry in arbitrary space-time dimensions on a square lattice. It has been recently shown that an extended version of the harmonic lattice is the discretized version of Lifshitz free scalar field theory\cite{MohammadiMozaffar:2017nri, He:2017wla}. By Lifshitz field theories we mean field theories which are invariant under the following scaling
$$t\to\lambda^z t\;\;\;\;\;,\;\;\;\;\;\vec{x}\to\lambda \vec{x},$$
in the massless limit. The extended version of this model is given by coupled harmonic oscillators with a generalized coupling term. The Hamiltonian is given by
\be\label{eq:LifH}
H=\sum_{n=1}^{N}\left[\frac{p_n^2}{2M}+\frac{M m^{2z}}{2}q_n^2+\frac{K}{2}\left(\sum_{k=0}^z(-1)^{z+k}{{z}\choose{k}} q_{n-1+k}\right)^2 \right].
\ee 
One can easily check that this Hamiltonian reduces to the well-known harmonic model for the case of $z=1$. One can also show that after a canonical transformation
$$(q_n,p_n)\to(\sqrt{MK}q_n,p_n/\sqrt{MK}),$$
this is a discretized version of a free Lifshitz theory on a square lattice with mass $m$ and lattice spacing $\epsilon=\sqrt{M/K}$,\footnote{In what follows for simplicity we choose $K=M=1$ without loss of generality.} i.e., the following action\cite{Alexandre:2011kr}
\bea\label{action}
S=\frac{1}{2}\int dt d\vec{x} \left[\dot{\phi}^2-\sum_{i=1}^{d}(\partial_i^z \phi)^2-m^{2z} \phi^2\right],
\eea
where $d$ is the number of spatial dimensions.\footnote{It is worth to note that some aspects of entanglement entropy of the same model has been previously studied in \cite{Nezhadhaghighi:2012vz, Nezhadhaghighi:2013mba, Nezhadhaghighi:2014pwa, Rajabpour:2014osa} for $0<z<1$. Also real-space vacuum entanglement of higher derivative scalar quantum field theories with specific values of $z$ has been computed in \cite{Kumar:2016ucp}.}

The Hamiltonian of this model \eqref{eq:LifH} on a (hyper)square lattice can be diagonalized (in arbitrary dimensions) in a standard way which we are not going to review here leading to the following dispersion relation \cite{MohammadiMozaffar:2017nri, He:2017wla}
\be
\omega^2_{\mathbf{k}}=m^{2z}+\sum_{i=1}^d\left(2\sin\frac{\pi k_{i}}{N_{x_i}}\right)^{2z},
\ee
where $\mathbf{k}=\{k_1,k_2,\cdots,k_d\}$ refers to the set of momentum components in all spatial directions and $k_{i}$ and $N_{x_i}$ refers to the momentum component and the number of sites in the $i$-th spatial direction respectively. 
It is worth to mention that as we have explained in \cite{MohammadiMozaffar:2017nri}, although in \eqref{eq:LifH} the dynamical critical exponent is an integer parameter, actually the resultant dispersion relation shows that exact analytic continuation to non integer values of $z$ is possible. At the moment we have no idea how to define the corresponding lattice Hamiltonian for generic $z$.

Implementing periodic boundary condition on all spatial directions leads to the following correlators
\begin{align}\label{eq:cor}
\begin{split}
\langle \phi_{\mathbf{i}}\phi_{\mathbf{j}} \rangle
&=\frac{1}{2}\prod_{r=1}^d \frac{1}{N_{x_r}}\sum_{k_r=0}^{N_{x_r}}\omega^{-1}_{\mathbf{k}}\coth\left(\frac{\omega_{\mathbf{k}}}{2T}\right)\cos\left(\frac{2\pi(i_r-j_r)k_r}{N_{x_r}}\right),
\\
\langle \pi_{\mathbf{i}}\pi_{\mathbf{j}} \rangle
&=\frac{1}{2}\prod_{r=1}^d \frac{1}{N_{x_r}}\sum_{k_r=0}^{N_{x_r}}\omega_{\mathbf{k}}\coth\left(\frac{\omega_{\mathbf{k}}}{2T}\right)\cos\left(\frac{2\pi(i_r-j_r)k_r}{N_{x_r}}\right),
\end{split}
\end{align}
for vacuum ($T=0$) and thermal states where $\mathbf{i}$ denotes a point on the (hyper)square lattice with coordinate $\{i_1,\cdots,i_d\}$.\footnote{These correlation functions are correct for both the vacuum state and the thermal state which we will consider in this paper and also for more general states at which $\langle a_\mathbf{i} a_\mathbf{j}\rangle=\langle a^\dagger_\mathbf{i} a^\dagger_\mathbf{j}\rangle=0 $.}
Also the corresponding correlators for Dirichlet boundary conditions at all boundaries is given by
\begin{align}\label{eq:corDir}
\begin{split}
\langle \phi_{\mathbf{i}}\phi_{\mathbf{j}} \rangle
&=\prod_{r=1}^d \frac{1}{N_{x_r}}\sum_{k_r=0}^{N_{x_r}}\tilde{\omega}^{-1}_{\mathbf{k}}\coth\left(\frac{\tilde{\omega}_{\mathbf{k}}}{2T}\right)\sin\left(\frac{\pi i_rk_r}{N_{x_r}}\right)\sin\left(\frac{\pi j_rk_r}{N_{x_r}}\right),
\\
\langle \pi_{\mathbf{i}}\pi_{\mathbf{j}} \rangle
&=\prod_{r=1}^d \frac{1}{N_{x_r}}\sum_{k_r=0}^{N_{x_r}}\tilde{\omega}_{\mathbf{k}}\coth\left(\frac{\tilde{\omega}_{\mathbf{k}}}{2T}\right)\sin\left(\frac{\pi i_rk_r}{N_{x_r}}\right)\sin\left(\frac{\pi j_rk_r}{N_{x_r}}\right),
\end{split}
\end{align}
where $\tilde{\omega}_{\mathbf{k}}=\omega_{\frac{\mathbf{k}}{2}}$. Once we have the correlators given in \eqref{eq:cor} and \eqref{eq:corDir}, one can work out the entanglement and Renyi entropies using Peschel's method \cite{Peschel}. To do so we define $X_{\mathbf{i}\mathbf{j}}=\langle \phi_\mathbf{i}\phi_\mathbf{j} \rangle$ and $P_{\mathbf{i}\mathbf{j}}=\langle \pi_\mathbf{i}\pi_\mathbf{j} \rangle$ where $\mathbf{i}, \mathbf{j}$ run over lattice sites in region $A$. The entanglement and Renyi entropies are given in terms of the eigenvalues of the operator $C=\sqrt{X\cdot P}$ which we denote by $\{\nu_k\}$ as the following
\begin{align}
S_A&=\sum_{k=1}^{N_A}\left[\left(\nu_k+\frac{1}{2}\right)\log\left(\nu_k+\frac{1}{2}\right)-\left(\nu_k-\frac{1}{2}\right)\log\left(\nu_k-\frac{1}{2}\right)\right],\label{EE}\\
S^{(n)}_A&=\frac{1}{n-1}\sum_{k=1}^{N_A}\log\left[\left(\nu_k+\frac{1}{2}\right)^n-\left(\nu_k-\frac{1}{2}\right)^n\right],\label{RE}
\end{align}
where $N_A$ is the number of sites in region $A$.

\begin{figure}
\begin{center}
\includegraphics[scale=0.35]{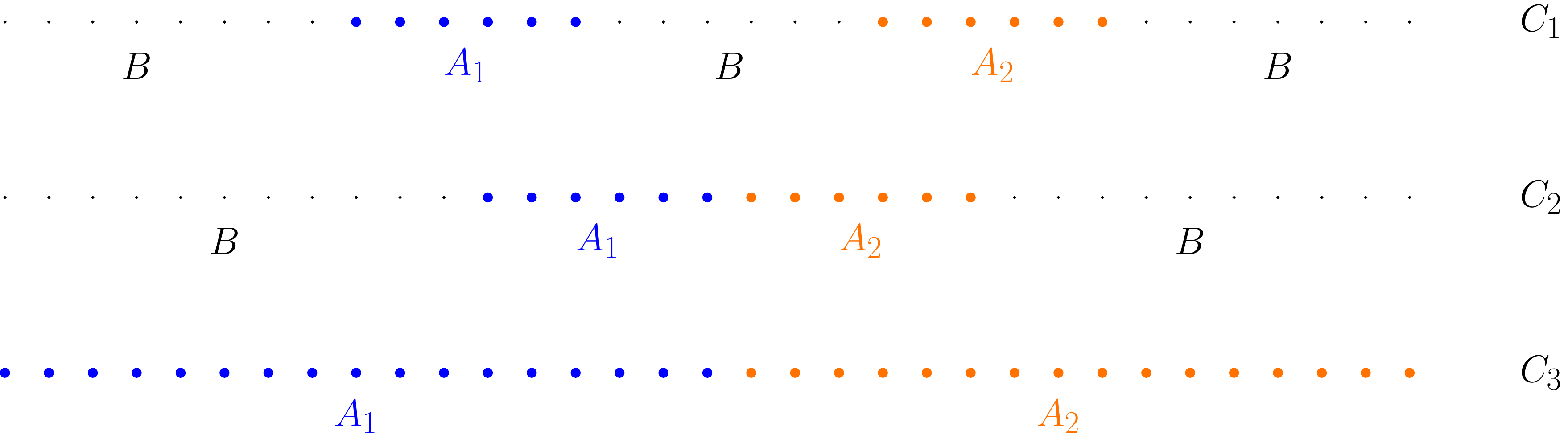}
\end{center}
\caption{Configurations we consider to studying negativity in $(1+1)$-dimensions. In $C_1$ we have two disjoint intervals while the complement $B$ is nonempty. In $C_2$ we have two adjacent intervals and the complement $B$ is nonempty again. In $C_3$ we have two adjacent intervals while the complement is empty. Note that in $C_1$ and $C_2$ the whole system can be either finite or infinite while in $C_3$ the system is finite.}
\label{fig:configurations1p1}
\end{figure}

\subsection{Partial Transpose and Logarithmic Negativity}
The key point which makes it possible to study logarithmic negativity in many-body bosonic systems is that taking the partial transpose with respect to a subregion $A_2$, is nothing but applying a time reversal operator on the momentum variables corresponding to the oscillators restricted to this region\cite{Audenaert}. This implies that a Gaussian state is transformed by the partial transposition to a Gaussian operator, and thus the correlator method to be applicable to compute negativity. 

To be more concrete we have to deal with the eigenvalues of the operator $C_{T_2}=\sqrt{X\cdot P^{T_2}}$ while $X$ is defined as previously. In order to define the operator $P^{T_2}$ we define
$$R_{A_2}=\mathrm{diag}\{1,1,\cdots,1,-1,-1,\cdots,-1\},$$
which is a square matrix of length $n_A$ and the $+1$'s correspond to the oscillators in $A_1$ and the $-1$'s elements correpond to the oscillators in $A_2$. Having this $P_{T_2}$ is defined as
$$P^{T_2}\equiv R_{A_2}\cdot P \cdot R_{A_2}.$$
The eigenvalues of $C_{T_2}$ which we denote by $\mu_i$'s are still positive definite but they do not need to satisfy in $\mu_i>\frac{1}{2}$. Having this we can find the moments of the reduced density matrix as
\be
\mathrm{Tr}\left(\rho^{T_2}_A\right)^n=\prod_{i=1}^{N_A}\left[\left(\mu_i+\frac{1}{2}\right)^n-\left(\mu_i-\frac{1}{2}\right)^n\right]^{-1},
\ee
thus the trace norm of the partial transposed reduced density matrix reads
\be
\left|\left|\rho^{T_2}_A\right|\right|=\prod_{i=1}^{N_A}\left[\left|\mu_i+\frac{1}{2}\right|-\left|\mu_i-\frac{1}{2}\right|\right]^{-1},
\ee
and the logarithmic negativity is given by
\be\label{lnega}
\mathcal{E}=\sum_{i=1}^{N_A}\log\left[\mathrm{max}\left(1,\frac{1}{2\mu_i}\right)\right].
\ee
In the following using the above expression we will study the logarithmic negativity in different set-ups.

\section{Logarithmic Negativity in (1+1)-dimensions}\label{sec:onepone}
In this section we first study logarithmic negativity in extended disjoint or adjacent subregions which we have plotted in figure \ref{fig:configurations1p1}. In the following we will consider the specific configurations known as $p$-alternating lattice configurations (see figure \ref{Fig:plattice}) which we are able to perform the calculation of logarithmic negativity analytically.

\begin{figure}
\begin{center}
\includegraphics[scale=0.27]{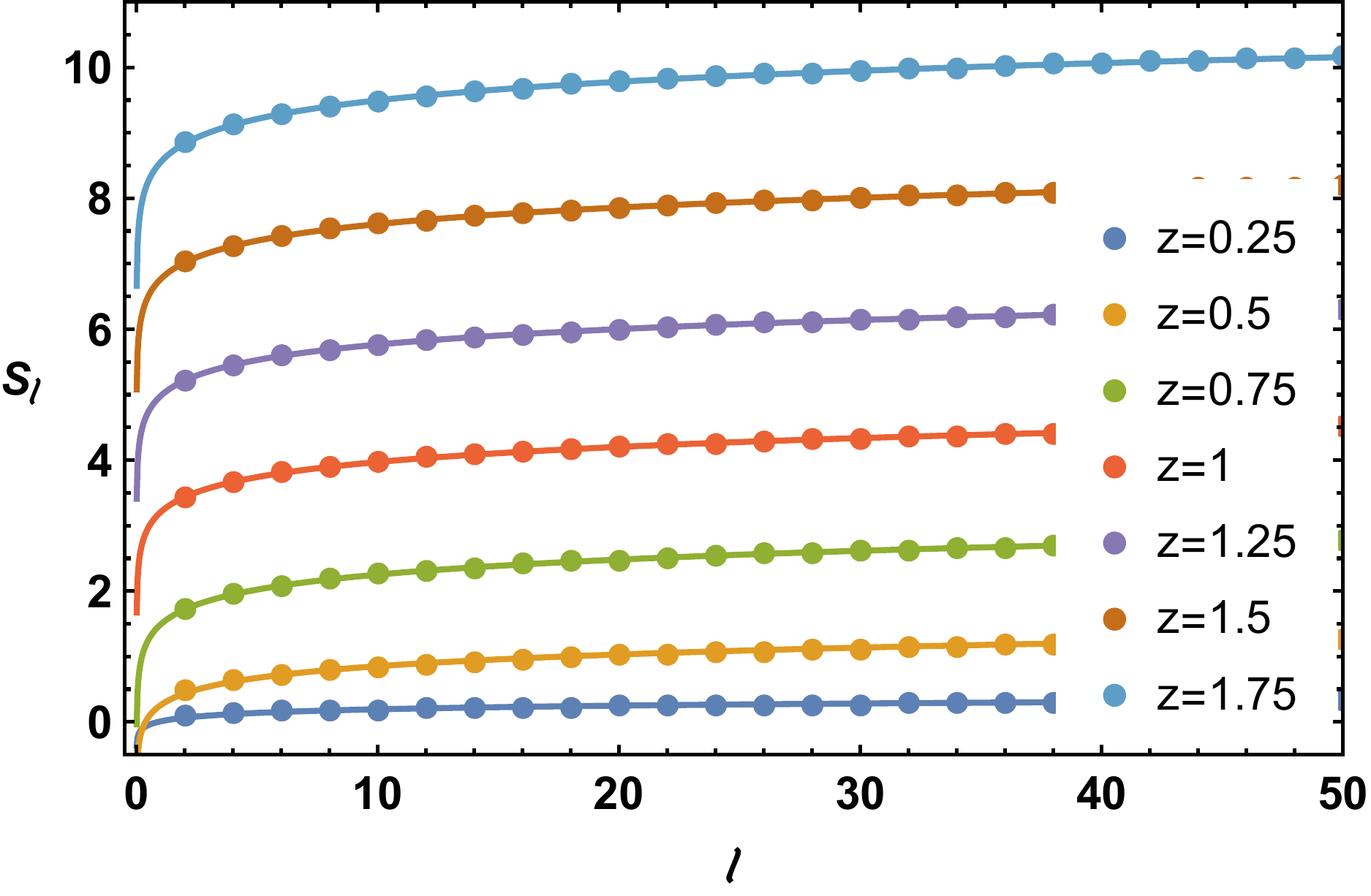}
\hspace{3mm}
\includegraphics[scale=0.27]{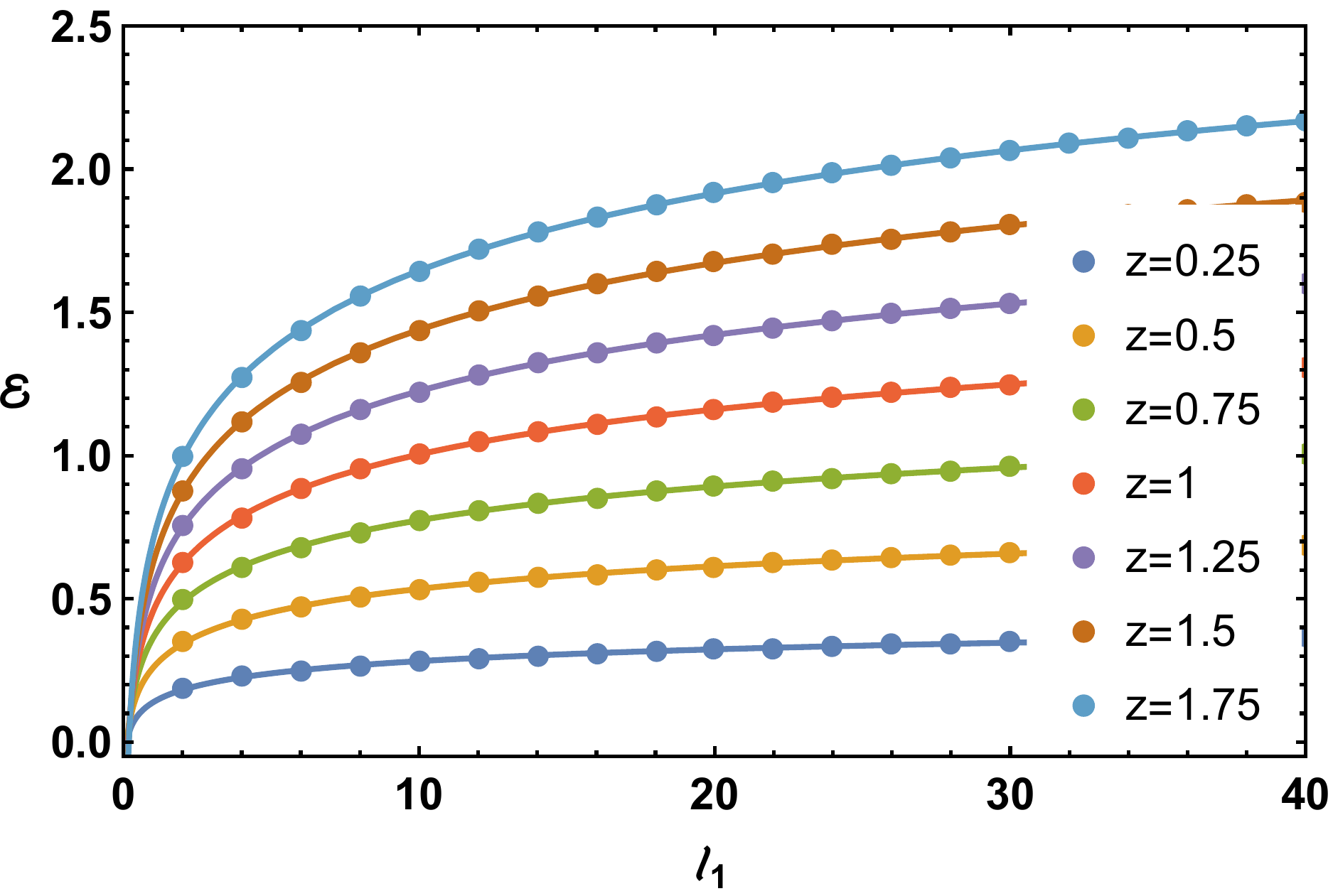}
\hspace{3mm}
\includegraphics[scale=0.27]{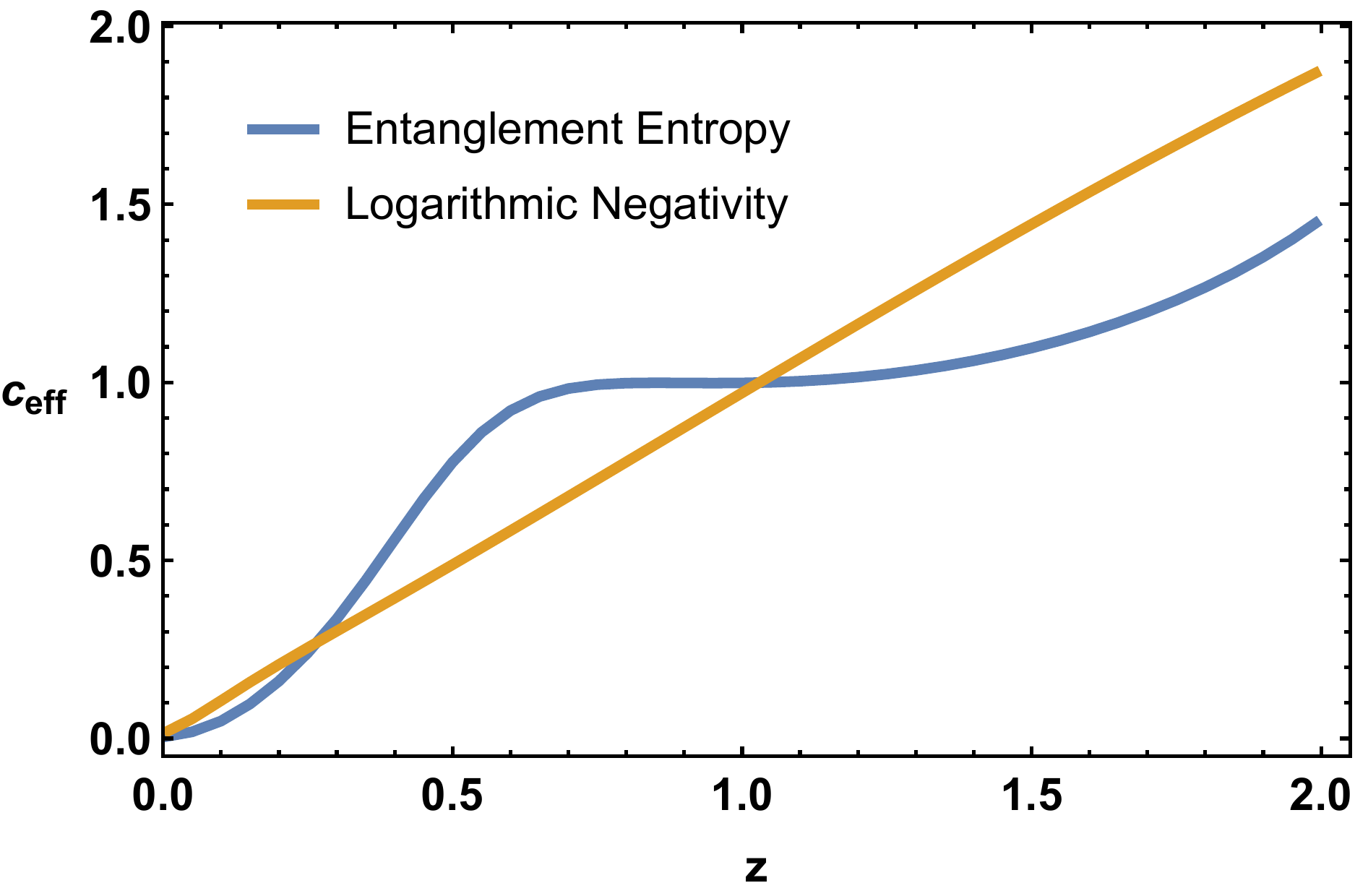}
\end{center}
\caption{Left panel shows entanglement entropy as a function of the length of the entangling region for $0<z<2$. The middle panel shows logarithmic negativity for configuration $C_2$ as a function of $\ell_1$. In this plot we have fixed $\ell_1+\ell_2=200$. The solid lines in both left and middle panels are fit functions similar to CFT analytic expression. The right panel shows the effective `c' read from entanglement entropy and logarithmic negativity. In all panels we have set the length of the chain to be $N_x=4000$ and $m=10^{-6}$ with periodic boundary condition.}
\label{fig:ceff}
\end{figure}

\subsection{Extended Subregions: Numerical Results}
Logarithmic negativity has been studied in the vacuum state of $(1+1)$-dimensional conformal field theories for adjacent and disjoint intervals and also in finite temperature states for different configurations using the Replica method. As the simplest example in the vacuum state of a CFT with an infinite spatial coordinate, logarithmic negativity for two adjacent intervals with length $\ell_1$ and $\ell_2$ up to a non-universal constant value is given by\cite{Calabrese:2012ew}
\be
\mathcal{E}=\frac{c}{4}\log\frac{\ell_1\ell_2}{\ell_1+\ell_2},
\ee
where $c$ is the central charge of the theory. This result is verified via the standard harmonic lattice model\footnote{By standard we mean $z=1$.} with periodic boundary condition in the conformal regime, i.e. $m N_x\ll1$ or even with Dirichlet boundary condition where we can set $m=0$. Also the well known result for entanglement entropy for single interval in the vacuum state of CFT is again up to a constant given by\cite{Calabrese:2004eu}
\be
S_{\ell}=\frac{c}{3}\log\frac{\ell}{\epsilon},
\ee
where $\ell$ is the length of the entangling region and $\epsilon$ is proportional to the inverse of the UV cutoff.

\begin{figure}
\begin{center}
\includegraphics[scale=0.26]{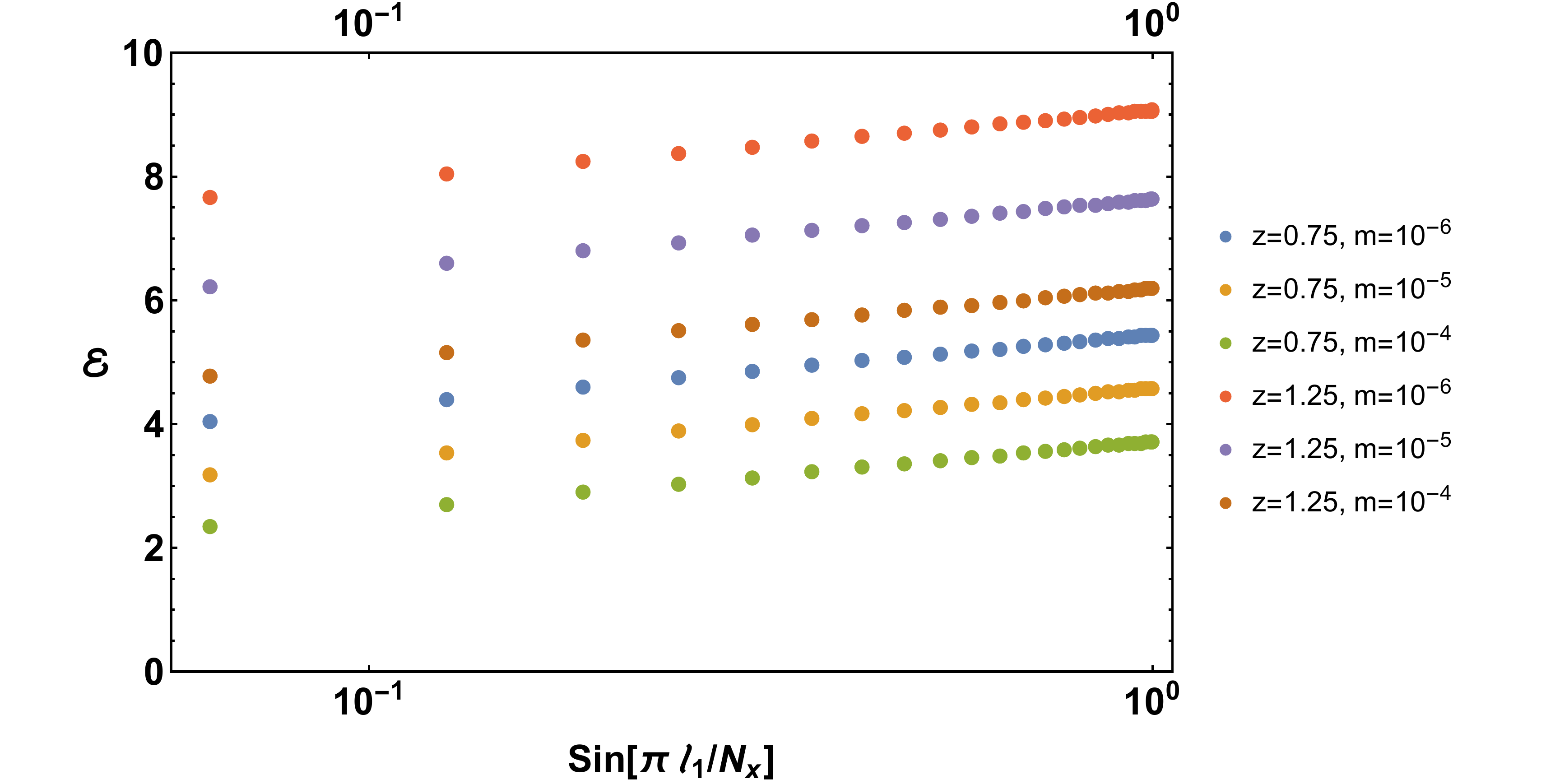}
\includegraphics[scale=0.68]{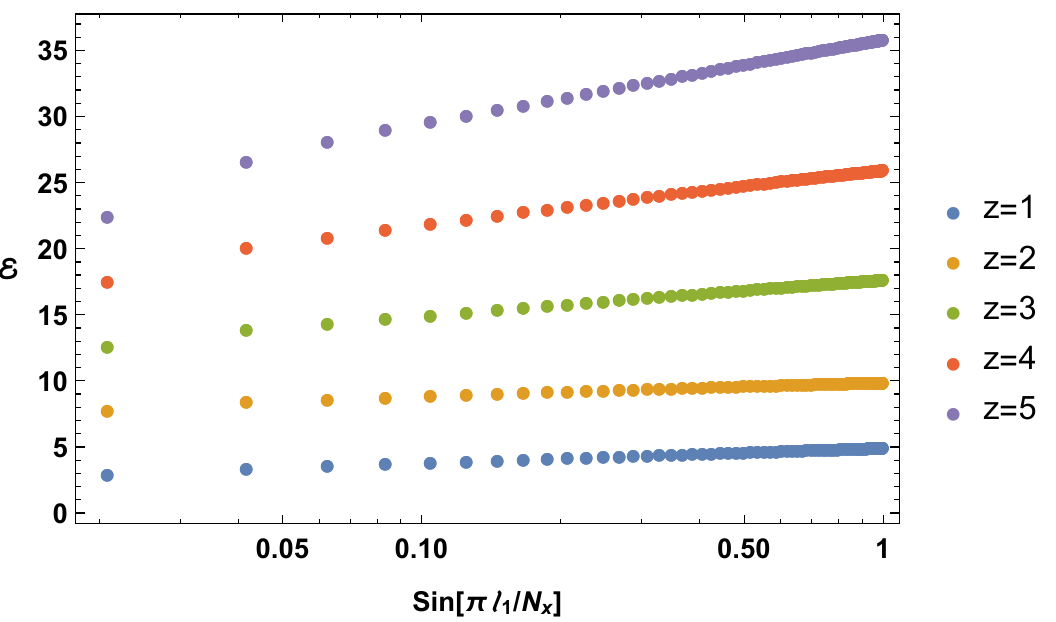}
\end{center}
\caption{\textit{Left}: Logarithmic negativity for the configuration $C_3$ as a function of $\ell_1$ in log-linear scale for a region with periodic BC for different values of $z$ and $m$. \textit{Right}: Same plot for larger values of $z$ for $m=10^{-4}$ and $N_x=100$.}
\label{fig:negativity-per1}
\end{figure}

Since we are not aware of an explicit way for calculation of Renyi entropies in field theories with Lifshitz symmetry, we proceed with numerically studying these models via the correlator method. In the following of this section we will study logarithmic negativity for different configurations shown in figure \ref{fig:configurations1p1} in both the vacuum state and in finite temperature states. 

As a warm up lets focus on the vacuum state and investigate whether these analytic expressions are valid for some range of the dynamical exponent or not. We find that these two expressions work excellently for $0<z<2$ for both cases. In figure \ref{fig:ceff} we have plotted some numerical data regarding to this range of parameters together with their fit functions with the same form as above letting $c$ and a constant to vary between different plots. We have considered the system to be very large ($N_x=4000$) in order to reproduce the CFT result at $z=1$. In the right panel we have plotted $c_{\mathrm{eff}}$ which is defined as
\be
\mathcal{E}=\frac{c_{\mathrm{eff}}}{4}\log\frac{\ell_1\ell_2}{\ell_1+\ell_2}+c_\mathcal{E}\;\;\;\;\;,\;\;\;\;\;
S_{\ell}=\frac{c_{\mathrm{eff}}}{3}\log \ell+c_S.
\ee
One can see that as expected at $z=1$ the lines intersect at $c_{\mathrm{eff}}=1$ which is showing the conformal case and there are three different regimes all over the window $0<z<2$ where these fit functions are valid. There is also another point which these two curves intersect at $(z\approx0.25,c_{\mathrm{eff}}\approx0.25)$.

As another check to verify the above analytic behavior one can check whether the mass which we have put as an IR cutoff on the periodic chain does effect the universal part or not. To see this we have plotted the negativity for two examples in this range which are  $z=0.75$ and $z=1.25$ for different values of parameter $m$ in the left panel of figure \ref{fig:negativity-per1}. One can see that the curves in the log-linear scale are parallel with each other showing that the effect of mass parameter is packed in the non-universal part and we should not worry about it. In the right panel of figure \ref{fig:negativity-per1} we consider the case with larger values of dynamical exponent where the validity of our fit functions is not obvious. According to this figure, once again the contribution due to a nonzero mass is not universal.

\begin{figure}
\begin{center}

\includegraphics[scale=0.26]{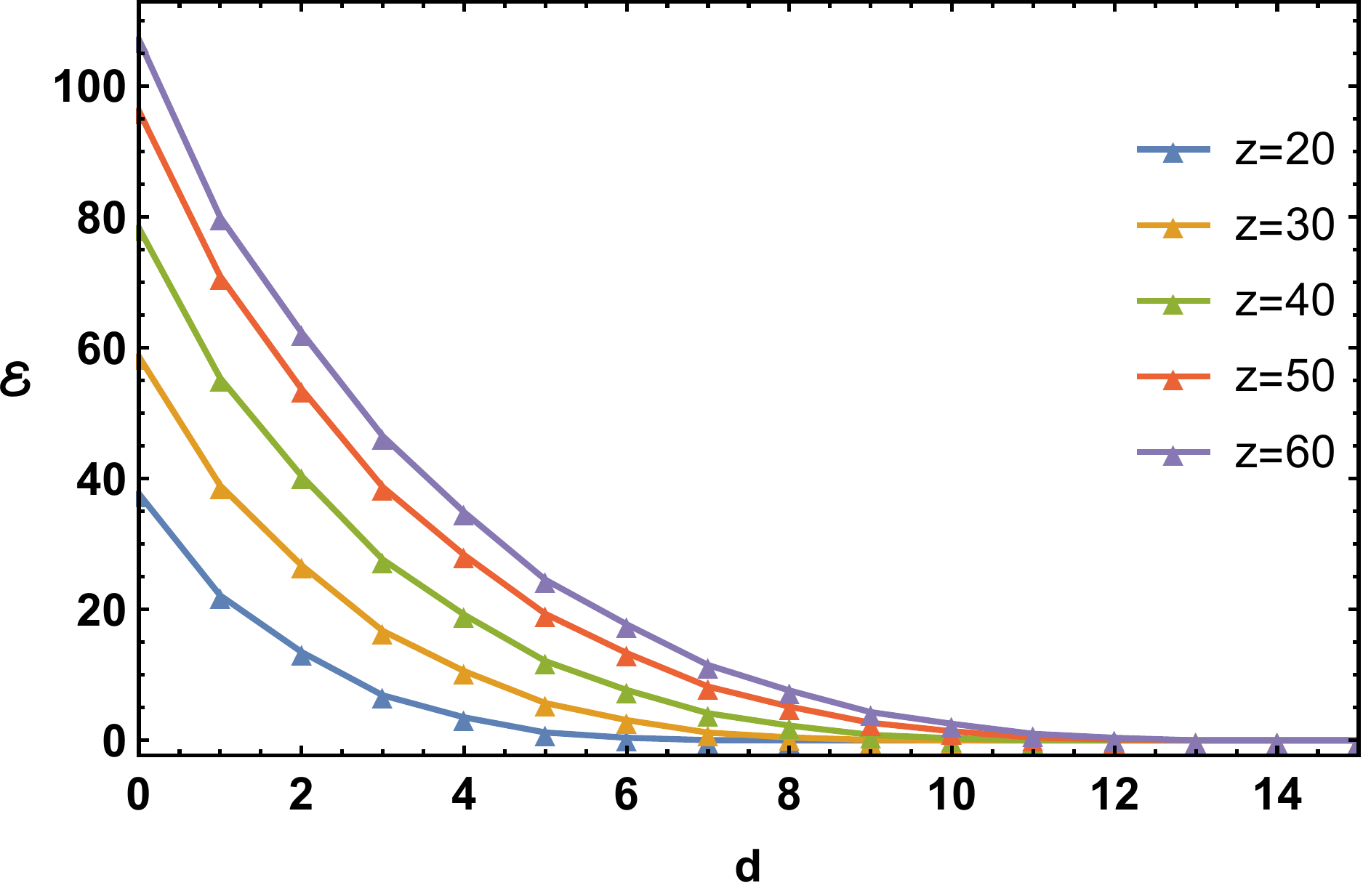}
\hspace{3mm}
\includegraphics[scale=0.26]{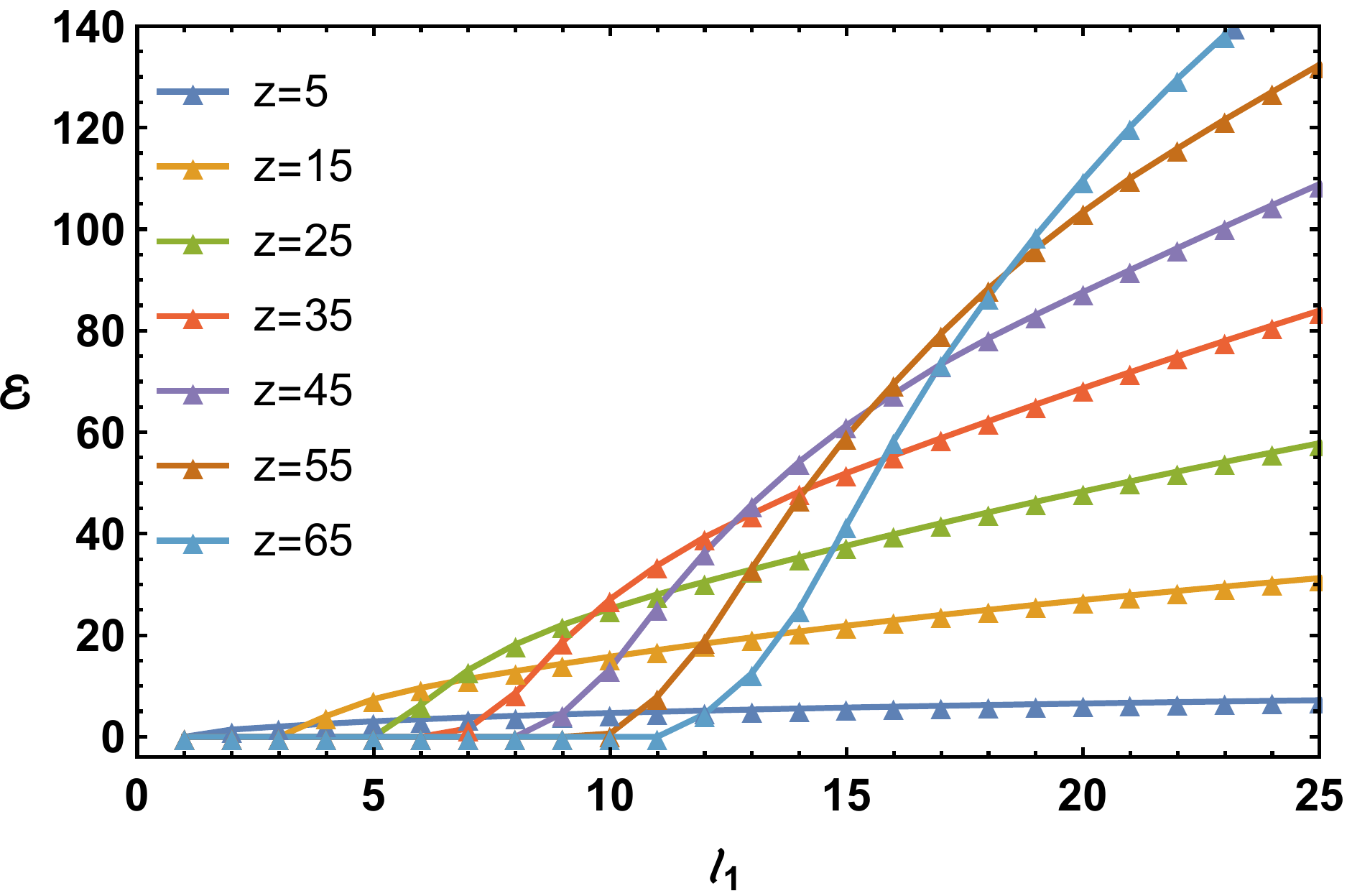}
\hspace{3mm}
\includegraphics[scale=0.55]{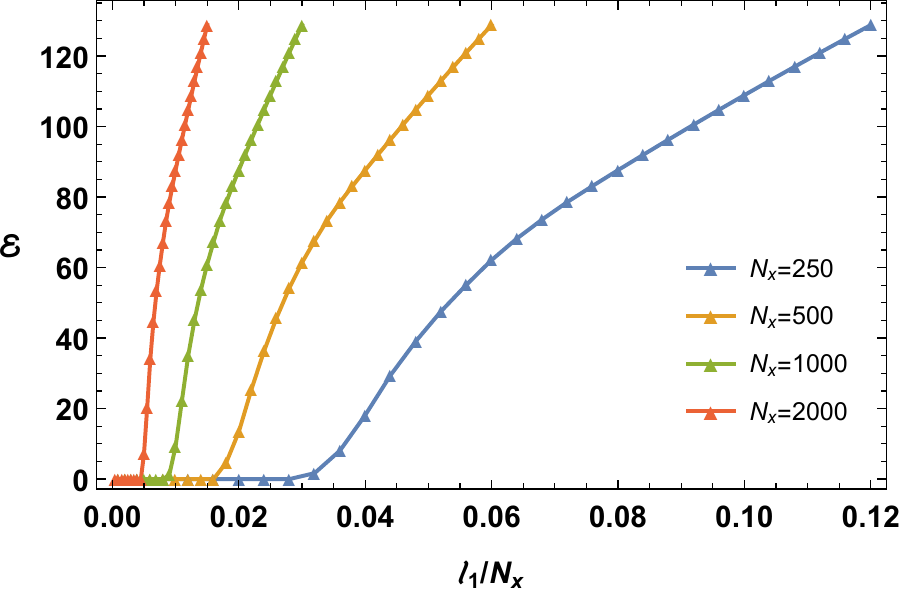}
\end{center}
\caption{The left panel shows logarithmic negativity for configuration $C_1$ versus the distance between $A_1$ and $A_2$ denoted by $d$. The parameters are $A_1=A_2=20$, $m=10^{-6}$ and $N_x=1000$. In the middle panel we have plotted the data regarding to configuration $C_2$ with the length of $A_1$ and $A_2$ being equal to each other on a chain of length $N_x=500$ and $m=10^{-4}$. In this figure as $z$ increases, the `initial sleep' region happens for larger subregions. The right panel shows the data for different chain lengths and $z=45$ indicating that `initial sleep' behavior disappears in the continuum limit. 
}
\label{fig:negativity-per}
\end{figure}

Figure \ref{fig:negativity-per} shows the logarithmic negativity for different values of dynamical exponent regarding to configurations $C_1$ and $C_2$. According to the left panel the logarithmic negativity decreases as we increase the separation between subsystems such that it is vanishing for $d\geq d_{\rm crit.}$. Also increasing the dynamical exponent logarithmic negativity increases due to the enhancement of correlations. In the middle panel we have plotted the data regarding to configuration $C_2$ for different values of dynamical exponent. According to this figure there is a regime where the logarithmic negativity vanishes for configurations smaller than a given length which depends on the value of the dynamical exponent. As we increase $z$ this `initial sleep' regime spreads over larger subregions. We believe that this peculiar behavior is a lattice effect. To see this we have plotted the right panel of figure \ref{fig:negativity-per}, which shows that increasing the total system size this phenomenon disappears in the continuum limit.\footnote{This `initial sleep' regime is similar to the `late birth of entanglement' phenomena which has been studied in \cite{Coser:2014gsa}. We would like to thank Andrea Coser for bringing our attention to this point.}

\subsection*{Thermal State}
We would now like to study logarithmic negativity in thermal states. Unlike entanglement entropy logarithmic negativity is a natural measure for these states and though there is no need to construct a non-pure density matrix via configurations like $C_1$ and $C_2$. Thus here we will naturally consider configuration $C_3$. It is worth to mention that using the replica trick people have studied logarithmic negativity in thermal states for $(1+1)$-dimensional CFTs \cite{Eisler2014,Calabrese:2014yza}. 
\begin{figure}
\begin{center}
\includegraphics[scale=0.32]{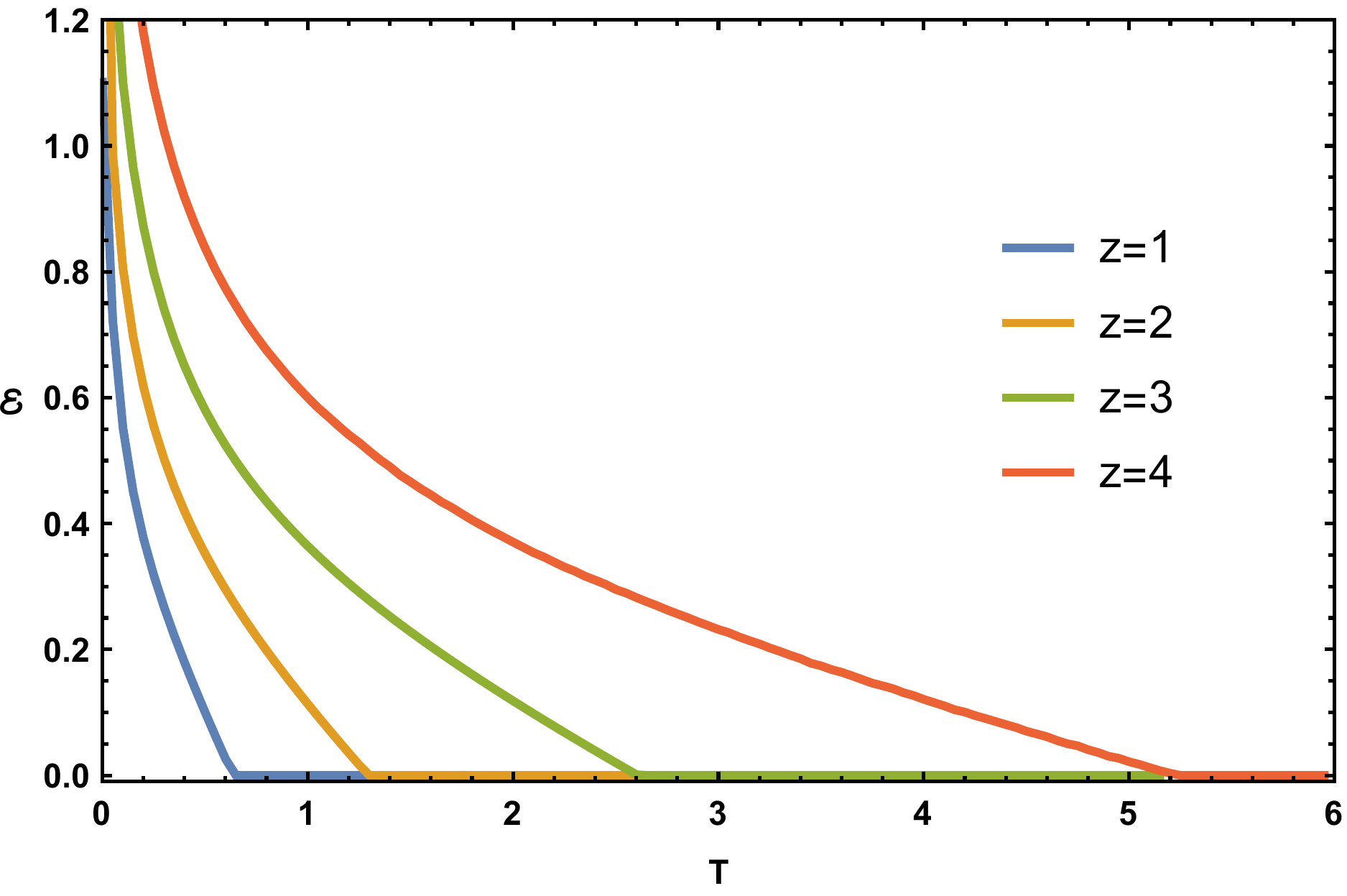}
\hspace{5mm}
\includegraphics[scale=0.68]{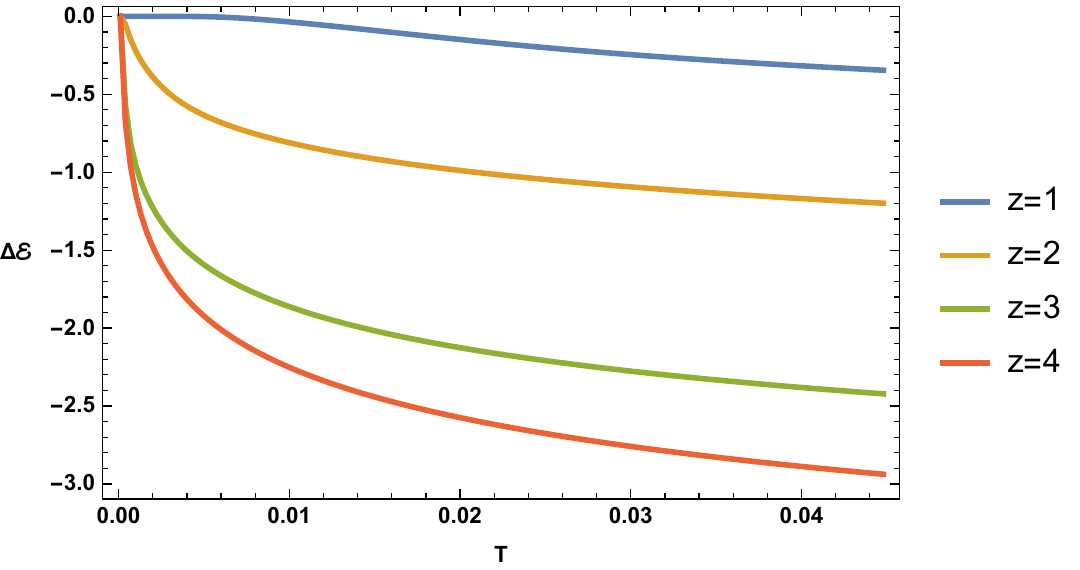}
\end{center}
\caption{Logarithmic negativity as a function of temperature for different values of $z$ for a region with Dirichlet BC. Here we set $m=0$ and $N_x=100$. \textit{Left}: $\mathcal{E}$ vanishes for $T>T_{\rm sd}$ due to lattice construction. \textit{Right}: In small temperature limit the magnitude of $\Delta \mathcal{E}$ increases for larger values of $z$.}
\label{fig:negfT}
\end{figure}

In the left panel of figure \ref{fig:negfT} we have plotted the numerical data for different values of dynamical exponent as a function of temperature. According to these curves the logarithmic negativity is a decreasing function of temperature in agreement with our expectation since in high temperature the classical correlations are dominant and a quantum system should crossover to a classical one. Once again increasing the dynamical exponent the correlation between subsystems increases and the logarithmic negativity increased. Note that for all values of $z$, the logarithmic negativity vanishes for $T>T_{\rm sd}$ which is a well known phenomena called sudden death entanglement \cite{yu:2009}. Actually as discussed in \cite{Anders:2008,Sherman,Calabrese:2012nk} this peculiar feature of logarithmic negativity is a numerical artifact due to the lattice construction and it is not relevant in the continuum limit where the system is described by the corresponding QFT. Thus, in order to avoid this lattice effect and to have clean results in the continuum limit, we consider the low-temperature regime as $T\ll \epsilon^{-z}$. The right panel of figure \ref{fig:negfT} shows the subtracted negativity, i.e., $\Delta \mathcal{E}\equiv \mathcal{E}(T)-\mathcal{E}(0)$, in this regime. Based on this figure, it is evident that in small temperature limit the magnitude of $\Delta \mathcal{E}$ increases while the dynamical exponent is increased. Also increasing the temperature makes the logarithmic negativity to decrease. Once again, this behavior is expected due to the diminution of quantum correlations in higher temperatures.

\subsection*{$z$-Dependence of Logarithmic Negativity}
In this section we would like to concentrate on the $z$-dependence of logarithmic negativity. From the numerical data reported up to now it is obvious that as the dynamical exponent increases, the value of logarithmic negativity also increases. We can understand this intuitively in terms of increasing number of correlated sites on the lattice in the Lifshitz harmonic lattice. Increasing behavior for other measures of entanglement in this model has been previously reported in \cite{MohammadiMozaffar:2017nri} and also some analytical results has been reported in \cite{He:2017wla} for specific highly symmetric configurations known as $p$-alternating lattice which we will discuss in the next part of this section. It has been shown that while increasing the dynamical exponent after an early quadratic-like growth, entanglement entropy soon reaches a regime which it grows linearly with $z$. This seems to be independent of the properties of the entangling region and also independent of space-time dimensions. We will show some analytical results for linear behavior in the next part of this section in $(1+1)$-dimensions and also some numerical results for higher dimensions in the following section.

To investigate the $z$-dependence of logarithmic negativity in $(1+1)$-dimensions, we have considered a $C_3$ configuration and studied entanglement negativity for different values of $z$. The corresponding numerical data are plotted for the vacuum state in the left panel of figure \ref{fig:negfT}. In this plot one can see that again for small values of the dynamical exponent entanglement negativity grows faster than linear with $z$ and for a total chain of length $N_x=100$ for $z\gtrsim 50$ it grows linearly with $z$. In the right panel of figure \ref{fig:negfT} we have plotted the data corresponding to thermal states. In this case the situation seems to be more complicated although as have shown in the figure again there seems to be a linear regime after a while of growth. As we will see in the next section in the low-temperature limit we can work out the $z$-dependence of the temperature correction to negativity for $p$-alternating lattice configurations which is a complicated function of the dynamical exponent.

\begin{figure}
\begin{center}
\includegraphics[scale=0.4]{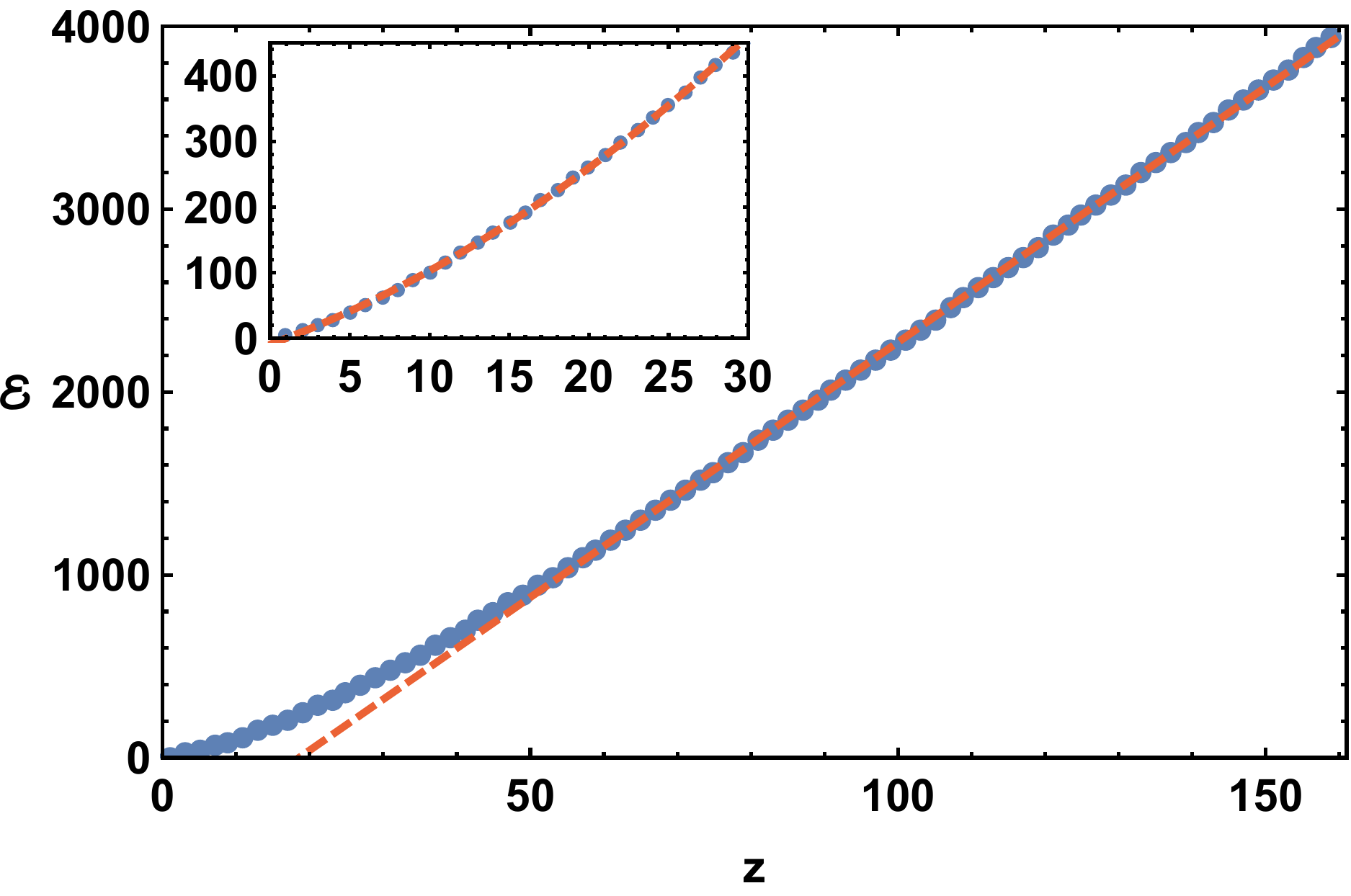}
\hspace{5mm}
\includegraphics[scale=0.38]{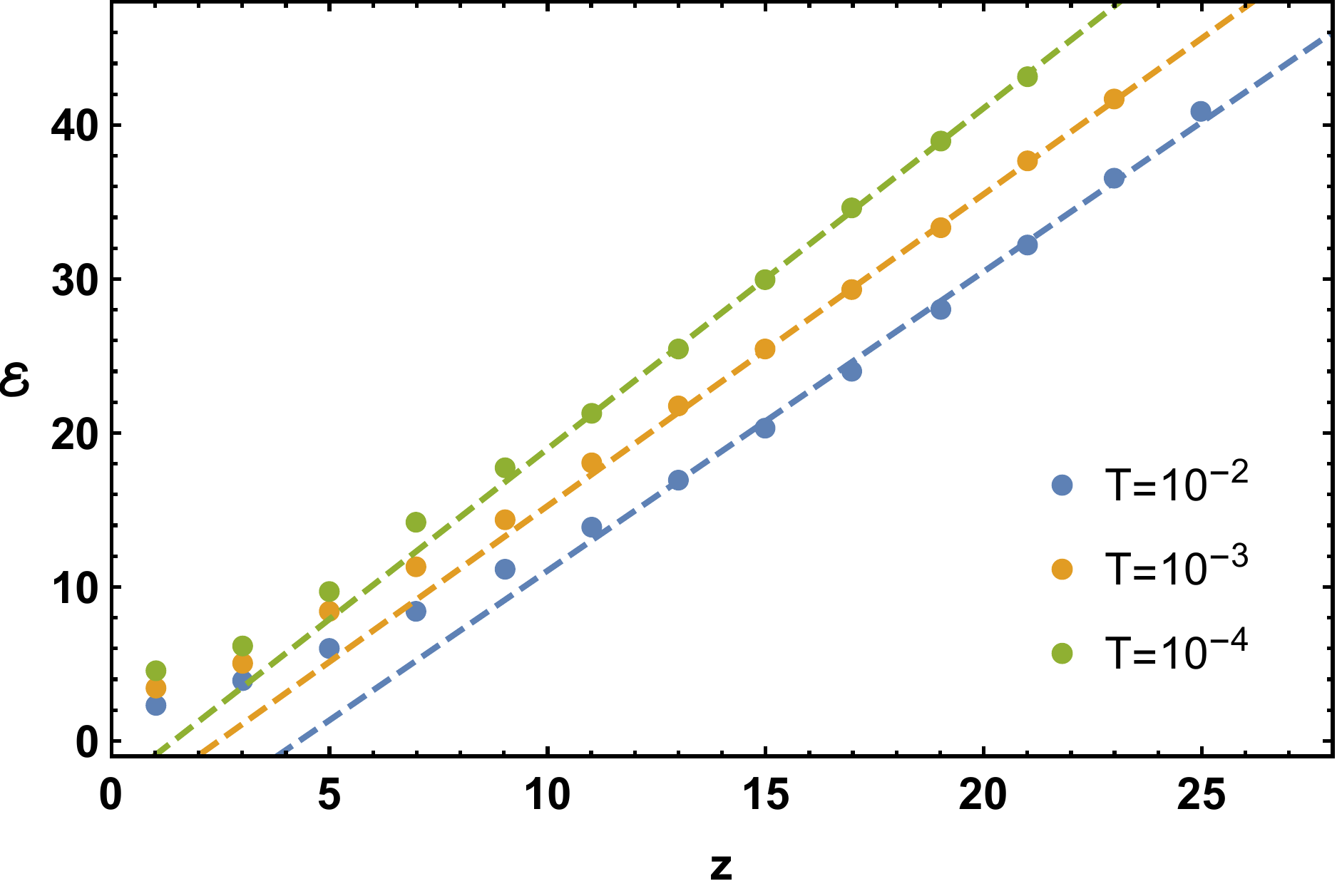}
\end{center}
\caption{Negativity as a function of $z$. The left panel shows negativity in the vacuum state for configuration $C_3$ on a chain of length $N_x=100$ with fixed $\ell_1=30$ and $m=10^{-4}$. The minor panel is the same curve focused on smaller values of $z$. The fit function used in this panel is of the form $a_0+a_1 z+a_2 z^2$. In the main panel the fit function is of the form $a_0+a_1 z$. The right panel is showing the data for thermal states. In this case the data correspond to  $C_3$ on a chain with $N_x=20$ with fixed $\ell_1=8$ and $m=10^{-5}$. The dashed lines refer to linear fits after excluding a few points corresponding small $z$.}
\label{fig:negZdep1p1}
\end{figure}

\subsection{$p$-alternating Lattice Subregions: Analytic Results}
In this section we study entanglement negativity in a specific configuration known as $p$-alternating sublattice \cite{He:2016ohr} on a lattice with periodic boundary condition. More precisely the $p$-alternating sublattice refers to a periodic decomposition of the lattice to $A$ and its complement $\bar{A}$ which the number of sites laying in $A$ is given by $N_A=N/p$ where $N$ is the total number of sites and $p\in \mathbb{Z}^+$. This kind of geometrical decomposition of the Hilbert space is not usually interesting in studying the entanglement structure in lattice models since it does not correspond to extended systems but it has a great advantage based on its highly symmetric structure. The advantage is that the correlator matrices introduced in \eqref{eq:cor} for such systems become \textit{circulant} matrices which their eigenvalues and eigenvectors are known analytically and thus the correlator method becomes analytically tractable. 

\begin{figure}
\begin{center}
\begin{tikzpicture}[>=latex, scale=.8]
\draw [color=gray!50](1, -1.732) arc (-60:360: 2cm and 2cm);
\draw (0, 0) coordinate (fov)
             +(0:2cm)  coordinate (A)  (fov)
           +(45:2cm)  coordinate (B)  (fov)
           +(90:2cm)  coordinate (C)  (fov)
           +(135:2cm)  coordinate (D)  (fov)
           +(180:2cm)  coordinate (E)  (fov)
           +(225:2cm)  coordinate (F)  (fov)
           +(270:2cm) coordinate (H) (fov)
           +(315:2cm) coordinate (G);
\fill[radius=2pt]
   (A)   node [fill, color=myblue, circle, inner sep=1pt ,minimum size=5pt]{}
   (B)   node [fill, color=myorange, circle, inner sep=1pt ,minimum size=5pt]{}
   (C)   node [fill, color=myblue, circle, inner sep=1pt ,minimum size=5pt]{}
   (D)   node [fill, color=myorange, circle, inner sep=1pt ,minimum size=5pt]{}
   (E)   node [fill, color=myblue, circle, inner sep=1pt ,minimum size=5pt]{}
   (F)   node [fill, color=myorange, circle, inner sep=1pt ,minimum size=5pt]{}
   (H)   node [fill, color=myblue, circle, inner sep=1pt ,minimum size=5pt]{}
   (G)   node [fill, color=myorange, circle, inner sep=1pt ,minimum size=5pt] {};
\draw [color=gray!50](9, -1.732) arc (-60:360: 2cm and 2cm);
\draw (8, 0) coordinate (fov)
           +(0:2cm)  coordinate (A)  (fov)
           +(30:2cm)  coordinate (B)  (fov)
           +(60:2cm)  coordinate (C)  (fov)
           +(90:2cm)  coordinate (D)  (fov)
           +(120:2cm)  coordinate (E)  (fov)
           +(150:2cm)  coordinate (F)  (fov)
           +(180:2cm) coordinate (G) (fov)
           +(210:2cm) coordinate (H) (fov)
           +(240:2cm) coordinate (II) (fov)
           +(270:2cm) coordinate (JJ) (fov)
           +(300:2cm) coordinate (KK) (fov)
           +(330:2cm) coordinate (LL);
\fill[radius=2pt]
   (A)   node [fill, color=gray, circle, inner sep=1pt ,minimum size=5pt]{}
   (B)   node [fill, color=myorange, circle, inner sep=1pt ,minimum size=5pt]{}
   (C)   node [fill, color=gray, circle, inner sep=1pt ,minimum size=5pt]{}
   (D)   node [fill, color=myblue, circle, inner sep=1pt ,minimum size=5pt]{}
   (E)   node [fill, color=gray, circle, inner sep=1pt ,minimum size=5pt]{}
   (F)   node [fill, color=myorange, circle, inner sep=1pt ,minimum size=5pt]{}
   (G)   node [fill, color=gray, circle, inner sep=1pt ,minimum size=5pt]{}
   (H)   node [fill, color=myblue, circle, inner sep=1pt ,minimum size=5pt] {}
   (II)   node [fill, color=gray, circle, inner sep=1pt ,minimum size=5pt] {}
   (JJ)   node [fill, color=myorange, circle, inner sep=1pt ,minimum size=5pt] {}
   (KK)   node [fill, color=gray, circle, inner sep=1pt ,minimum size=5pt] {}
   (LL)   node [fill, color=myblue, circle, inner sep=1pt ,minimum size=5pt] {};
\end{tikzpicture} 
\caption{$p$-alternating periodic sublattice configurations. The left panel shows the case of $N=8$ while $p=1$ and $p'=2$. In this case the region $B$ is empty and the sites in $A_1$ are shown by blue and the sites in $A_2$ with orange. In the right panel we have shown the case of $N=12$, where $p=p'=2$. Again the sites in $A_1$ are shown by blue, those in $A_2$ with orange and those in $B$ by gray.}
\label{Fig:plattice}
\end{center}
\end{figure}
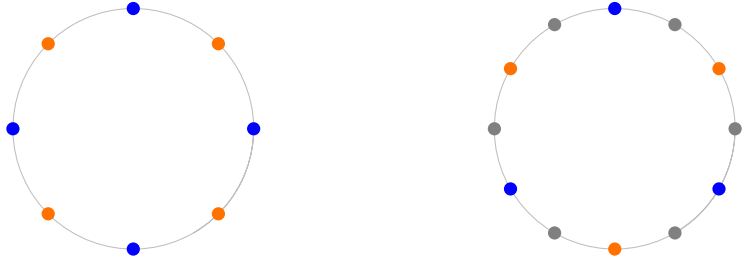

Previously people have used this method to study entanglement entropy of such systems in harmonic lattice while the system is in an eigenstate of the total Hamiltonian or in a thermal state \cite{He:2016ohr} (see also \cite{He:2017wla} for a similar study in Lifshitz harmonic lattice). Here we introduce a similar setup in order to analytically study negativity in such a configuration. As we mentioned before in order to end up with a mixed state and study logarithmic negativity, one possible way is to decompose the region $A=A_1\cup A_2$ and trace out the complement of $A$. A simple analysis shows that by choosing $N_A=p'N_{A_1}$ with $p'=2$ the partial transpose of the momentum correlation matrix, i.e. $P^{T_2}=R_{A_2}\cdot P\cdot R_{A_2}$ becomes a circulant matrix and one can find the analytic expression for the spectrum of $\rho_A^{T_2}$. Also in what follows we always consider the case where $N_A$ is an even number to more simplify the expressions. This choice for the decomposition of $A$ into sublattices is depicted for two examples in figure \ref{Fig:plattice}. In this case the $R_{A_2}$ which is a $N_A\times N_A$ matrix becomes
\bea
R_{A_2}=
\begin{pmatrix}
            1 & 0 & 0 & \cdots &0   \\
           0 &-1&0&\cdots &0 \\
             0&0&1&\cdots &0 \\
            \vdots &\vdots &\vdots &\ddots &\vdots \\
             0&0&0&\cdots &-1 \\
             \end{pmatrix},\;\;\; \left(R_{A_2}\right) _{IK}=\delta_{IK}e^{i \pi I}.
\eea

\begin{figure}
\begin{center}
\includegraphics[scale=0.5]{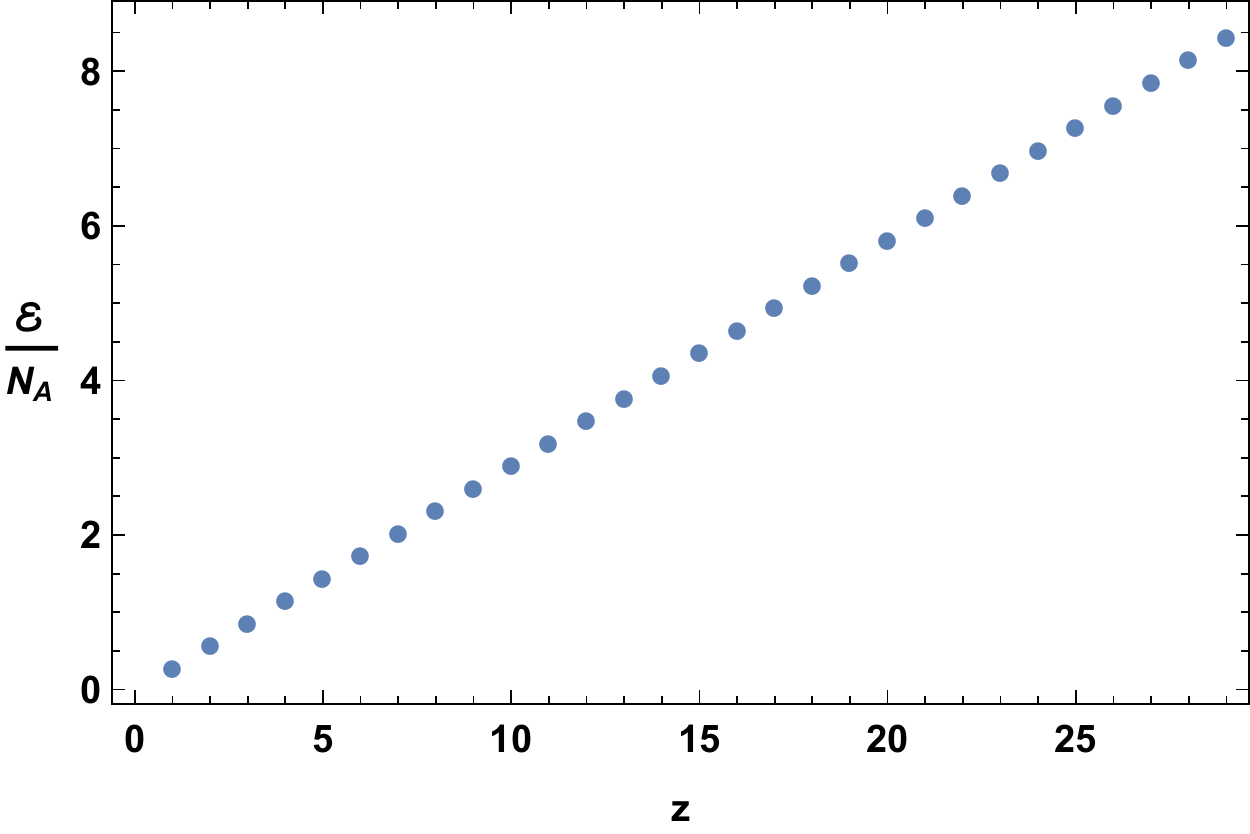}
\hspace{5mm}
\includegraphics[scale=0.93]{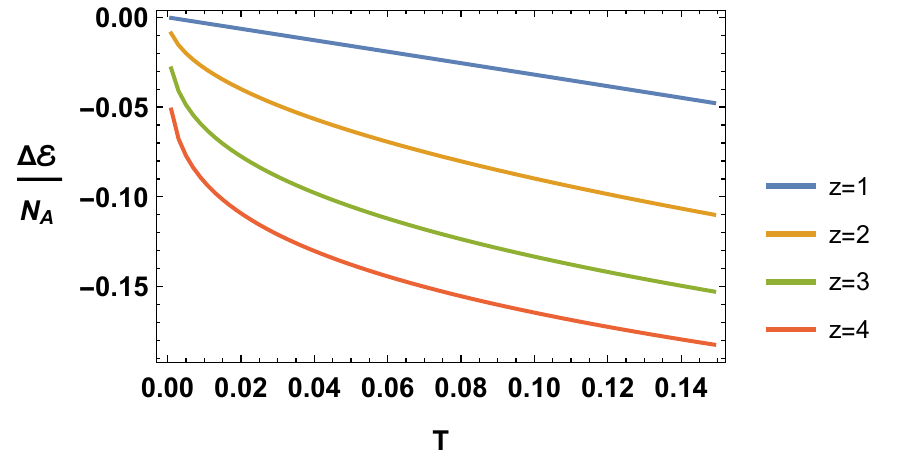}
\end{center}
\caption{Logarithmic negativity for a $p$-alternating sublattice configuration for $p=1$ and $m=0$ plotted by analytic results. The left panel shows negativity as a function of $z$ and the right panel is showing the results for thermal states for different values of dynamical exponents.}
\label{fig:analytic}
\end{figure}

Now using Eq.\eqref{eq:cor} the corresponding correlators are given as follows
\bea
X_{ij}&=&\frac{1}{2N}\sum_{k=0}^{N-1}\omega_k^{-1}\coth \frac{ \omega_k}{2T}\cos\frac{2\pi(i-j)k}{N_A},\nonumber\\
\left(P^{T_2}\right)_{ij}&=&\frac{e^{i\pi(i-j)}}{2N}\sum_{k=0}^{N-1}\omega_k\coth \frac{ \omega_k}{2T}\cos\frac{2\pi(i-j)k}{N_A},
\eea
with $i,j=0,\cdots, N_A-1$. The eigenvalues of a generic $N_A\times N_A$ circulant matrix parametrized as
$$C={\rm circ}(c_0, c_1, \cdots, c_{N_A-1}),$$
are given by
\bea
\mu_l=\sum_{k=0}^{N_A-1}c_k e^{-\frac{2\pi i kl}{N_A}}\;\;\;\;\;,\;\;\;\;\;l=0, \cdots, N_A-1.
\eea
Using the above relation since $X$ and $P^{T_2}$ are circulant matrices, it is an easy task to find the corresponding eigenvalues for $\sqrt{X\cdot P^{T_2}}$. Denoting the corresponding eigenvalues for $X$ and $P^{T_2}$ by $\mu_l^{(X)}$ and $\mu_{l}^{(P^{T_2})}$ the eigenvalues for $\sqrt{X\cdot P^{T_2}}$ are given by $\mu_l=\sqrt{\mu_l^{(X)}\cdot \mu_{l-\frac{N_A}{2}}^{(P^{T_2})}}$ which can be simplified as follows
\begin{align}
\begin{split}
\label{mutemp}
\mu_l&=\frac{1}{4p}\Bigg[\sum_{j=0}^{p-1}\left(\frac{\coth\frac{ \omega_{jN_A+l}}{2T}}{\omega_{jN_A+l}}+\frac{\coth\frac{ \omega_{(j+1)N_A-l}}{2T}}{\omega_{(j+1)N_A-l}}\right)\times
\\&\hspace{2.5cm}
\sum_{k=0}^{p-1}\left(\omega_{(k-\frac{1}{2})N_A+l}\coth\frac{ \omega_{(k-\frac{1}{2})N_A+l}}{2T}+\omega_{(k+\frac{3}{2})N_A-l}\coth\frac{ \omega_{(k+\frac{3}{2})N_A-l}}{2T}\right)\Bigg]^{1/2}.
\end{split}
\end{align}
Now we are equipped with all we need to calculate the logarithmic negativity for this configuration using equation \eqref{lnega}. The above expression has a much simpler form in the zero temperature limit as
\bea
\mu_l=\frac{1}{2p}\left(\sum_{j,k=0}^{p-1}\frac{\omega_{(k-\frac{1}{2})N_A+l}}{\omega_{jN_A+l}}\right)^{1/2}.
\eea
Now restricting to the continuum limit, i.e. $N, N_A\gg 1$, with $\frac{N}{N_A}=p$ fixed, the above eigenvalues can be rewritten as follows
\bea\label{coneigen}
\mu(x)=\frac{1}{2p}\left(\sum_{j,k=0}^{p-1}\frac{\omega(x+\frac{k-1/2}{p})}{\omega(x+j/p)}\right)^{1/2},\;\;\;\;\omega(x)^2=m^{2z}+(2\sin \pi x)^{2z}
\eea
where $0\leq x\equiv \frac{l}{N}<\frac{1}{p}$. In this case the negativity becomes
\be\label{connega}
\mathcal{E}=-N\int_0^{\frac{1}{p}}dx\log \left(\left|\mu(x)+\frac{1}{2}\right|-\left|\mu(x)-\frac{1}{2}\right|\right).
\ee
To simplify the analysis and illustrate this behavior, let us consider the
massless regime with $p=1$.  In this case using Eq.\eqref{coneigen} the corresponding eigenvalues is given by
\bea
\mu(x)=\frac{1}{2}\sqrt{\frac{\omega(x-\frac{1}{2})}{\omega(x)}}=\frac{1}{2}\cot^{\frac{z}{2}}(\pi x).
\eea
Now Eq.\eqref{connega} becomes
\be
\mathcal{E}=-N\int_{0}^1dx\log \left(\frac{1}{2}\left|\cot^{\frac{z}{2}}\pi x+1\right|-\frac{1}{2}\left|\cot^{\frac{z}{2}}\pi x-1\right|\right)=N\frac{z}{2}\int_{\frac{1}{4}}^{\frac{3}{4}}dx\log\left|\tan \pi x\right|.
\ee
Finally noting that for $p=1$ we have $N=N_A$ one finds
\be\label{negap1}
\frac{\mathcal{E}}{N_A}=\frac{G_c}{\pi}z,
\ee
where $G_c$ is the Catalan's constant which its numerical value is approximately $G_c\approx 0.916$. In the left panel of figure \ref{fig:analytic} we have plotted this linear behavior.
An interesting feature of the above result is that the logarithmic negativity is proportional to the volume of the corresponding subregion, i.e., $N_A$, and also it depends  linearly on the dynamical exponent. Also note that this simple result which is valid for $p=1$ shows an important feature of the negativity for quantum systems described by a pure state. Actually the $p$-alternating sublattice that we consider in zero temperature limit for $p=1$ is described by a pure state. It is well-known that logarithmic negativity for pure states is given by Renyi entropy for $n=\frac{1}{2}$, i.e., $\mathcal{E}=S_A^{(1/2)}$ \cite{Calabrese:2012ew}. In order to check this property in our setup note that at zero temperature when the $p$-alternating sublattice is divided into two parts the corresponding eigenvalues for $\sqrt{X\cdot P}$ for a massless scalar in the continuum limit is given by\footnote{Note that considering the case of entanglement and Renyi entropies for a two-partite system we should consider $p=2$ in agreement with \cite{He:2017wla}. On the other hand for logarithmic negativity because we always set $p'=2$, one must consider $p=1$ to have a two-patite system.}\cite{He:2017wla}
\bea
\nu(x)=\frac{1}{4}\left(\sum_{j,k=0}^{1}\frac{\omega(x+\frac{k}{p})}{\omega(x+\frac{j}{p})}\right)^{1/2}=\frac{1}{4}\left|\tan^{\frac{z}{2}} \pi x+\cot^{\frac{z}{2}} \pi x\right|.
\eea
Thus, by employing Eq.\eqref{RE} for $n=\frac{1}{2}$, we find 
\bea
S_A^{(1/2)}&=&N\int_{0}^{\frac{1}{2}}dx\log\left(\sqrt{\nu(x)+\frac{1}{2}}-\sqrt{\nu(x)-\frac{1}{2}}\right)\nonumber\\
&=&\frac{Nz}{4}\left(\int_{0}^{\frac{1}{4}}dx \log \tan \pi x-\int_{\frac{1}{4}}^{\frac{1}{2}}dx \log \tan \pi x\right).
\eea
Doing the above integral and noting that $N=2N_A$ the above result can be written as 
\bea
\frac{S_A^{(1/2)}}{N_A}=\frac{G_c}{\pi}z,
\eea
in agreement with Eq.\eqref{negap1}.

As we mentioned before the logarithmic negativity seems to be a suitable quantum  measure for mixed states, e.g., thermal states. So it is interesting to extend our analysis to the case of nonzero temperature and investigate the thermal corrections to logarithmic negativity. Once again we consider the configuration with $p=1$. In this case using equation \eqref{mutemp} the corresponding eigenvalues in the continuum limit is given by
\bea
\mu(x)=\frac{1}{2}\sqrt{\frac{\omega(x-\frac{1}{2})}{\omega(x)}\coth\frac{\omega(x)}{2T}\coth\frac{\omega(x-\frac{1}{2})}{2T}}.
\eea
Substituting the above expression in equation \eqref{connega} and performing similar steps as in the zero temperature case, one can find thermal corrections to logarithmic negativity as follows
\bea
\frac{\Delta\mathcal{E}}{N_A}=\frac{1}{2}\int_{\frac{1}{4}}^{\frac{3}{4}}dx \log\left(\tanh\frac{\omega(x)}{2T}\tanh \frac{\omega(x-\frac{1}{2})}{2T}\right),
\eea
where again we have defined $\Delta\mathcal{E}=\mathcal{E}(T)-\mathcal{E}(0)$. Focusing on the low-temperature limit we have
\bea
\frac{\Delta\mathcal{E}}{N_A}=-\int_{\frac{1}{4}}^{\frac{3}{4}}dx \left[e^{-\frac{\omega(x)}{T}}+e^{-\frac{\omega(x-\frac{1}{2})}{T}}+\mathcal{O}\left(e^{-\frac{3\omega(x)}{T}},e^{-\frac{3\omega(x-\frac{1}{2})}{T}}\right)\right],
\eea
where $\omega(x)=(2\sin \pi x)^z$. The above result is illustrated in the left panel of figure \ref{fig:analytic} for different values of dynamical exponent. According to this figure for $z=1$ we have a linear behavior for $\Delta\mathcal{E}$ but for larger values of $z$, the linear behavior is reached at higher temperatures. 

\section{Logarithmic Negativity in (2+1)-dimensions}\label{sec:twopone}
In this section we focus on $(2+1)$-dimensions. In this case beside what we have investigated in $(1+1)$-dimensions, several new questions may be asked regarding to the possibility of shape dependence of subregions in study. Logarithmic negativity has been previously studied for bosonic systems in $(2+1)$-dimensions for the $z=1$ case in \cite{DeNobili:2016nmj}. In this section we will study logarithmic negativity for generic $z$ and in some cases compare the results with those available in the literature for the Lorentzian case.

\begin{center}
\begin{figure}
\includegraphics[scale=0.35]{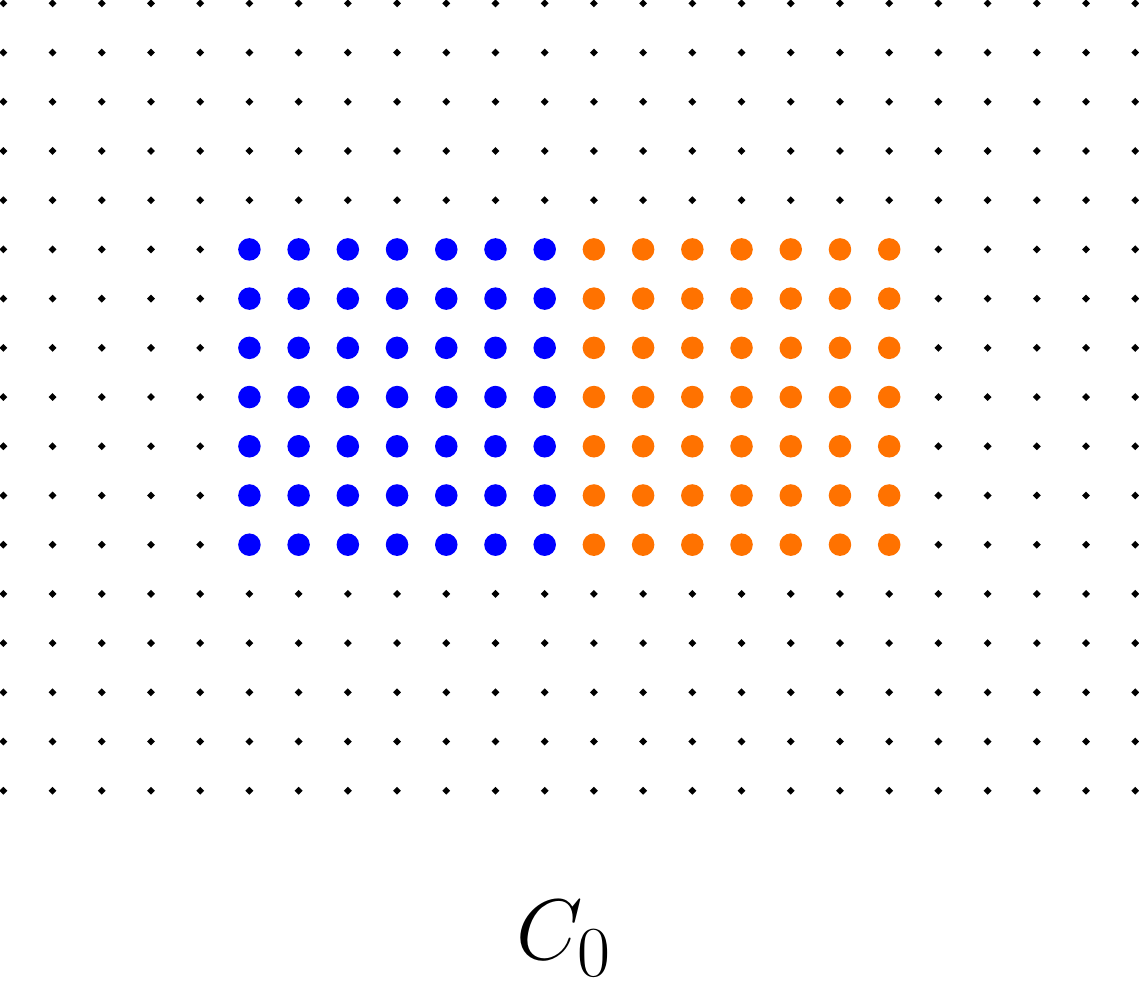}
\hspace{3mm}
\includegraphics[scale=0.35]{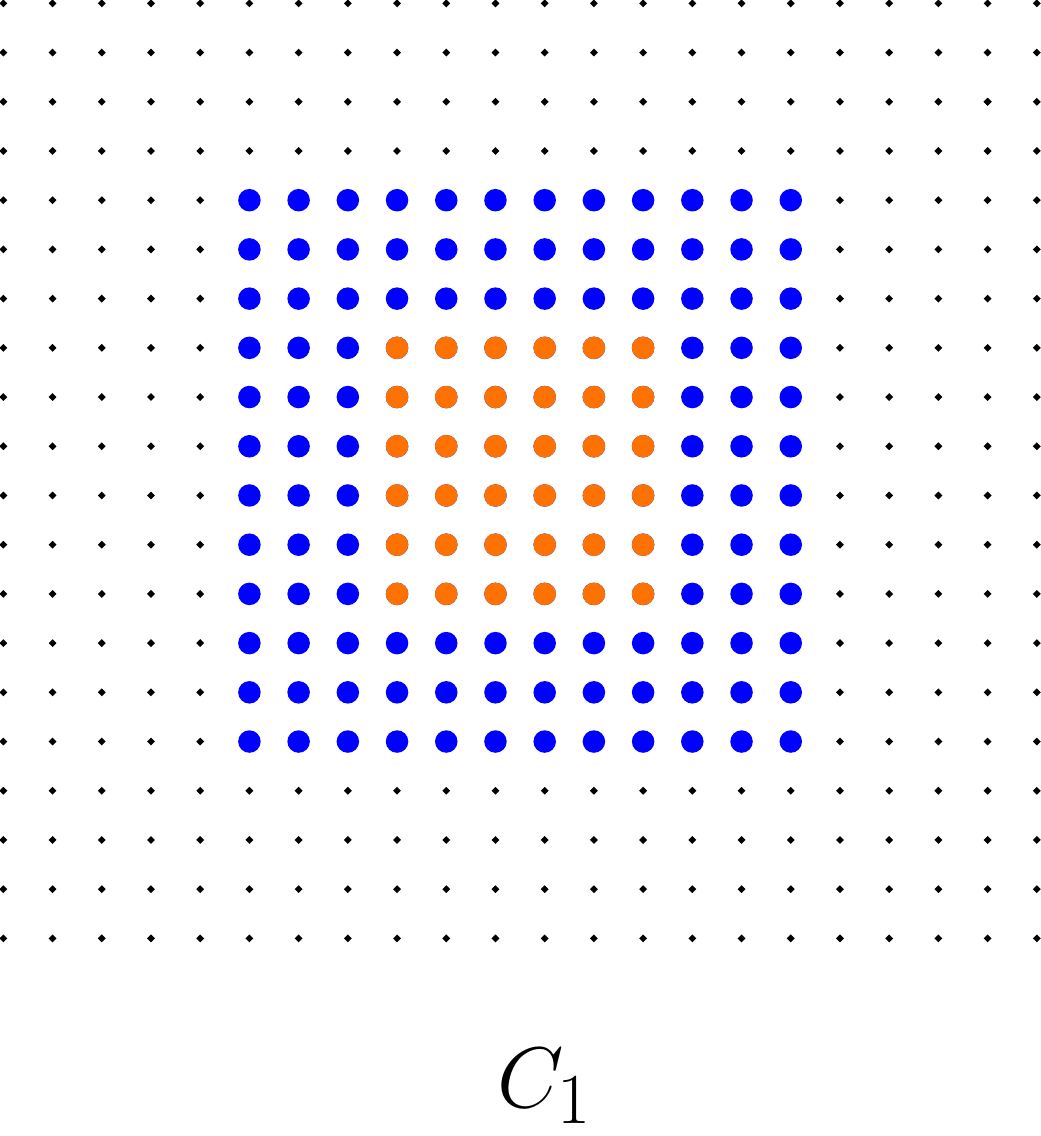}
\hspace{3mm}
\includegraphics[scale=0.35]{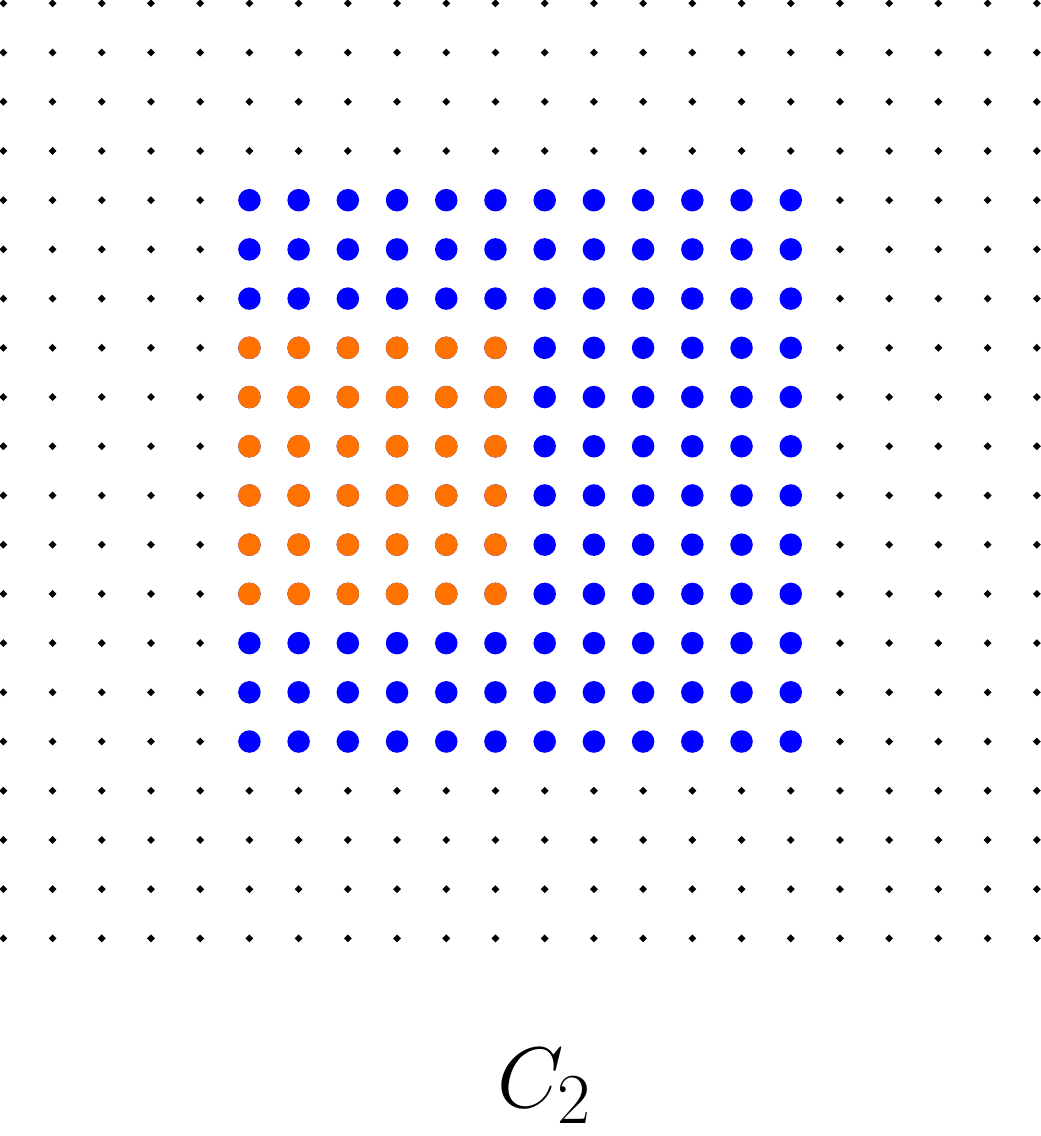}
\hspace{3mm}
\includegraphics[scale=0.35]{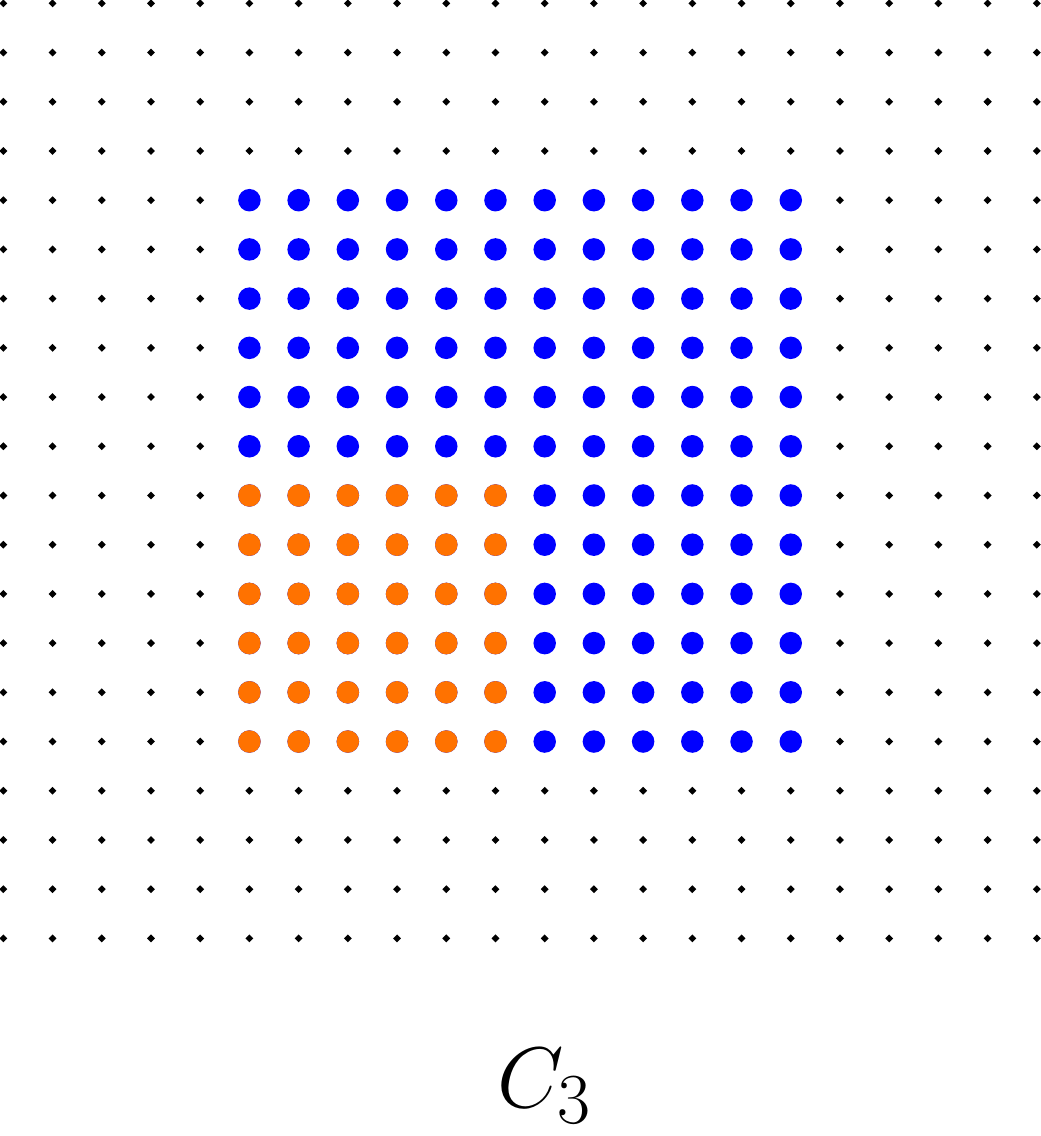}
\caption{Configurations we consider to studying negativity in $(2+1)$-dimensions. In all these configurations the colored (blue and orange) sites are those in region $A$ and the black sites lay in the complement. The blue sites refer to $A_1$ and the orange sites to $A_2$.}
\label{fig:configurations}
\end{figure}
\end{center}

Before starting our detailed analysis we introduce the configurations which we have considered in this section which are illustrated in figure \ref{fig:configurations}. The configurations are showing the generic case where the complement of region $A$ in not empty. The blue sites refer to $A_1$ and the orange ones refer to $A_2$. In the very left configuration which we denote by $C_0$, our parametrization is such that region $A$ is a rectangle with sides $2\ell_x$ and $\ell_y$, i.e. $A_1$ and $A_2$ are both  rectangles with sides $\ell_x$ and $\ell_y$. In the other three configurations denoted by $C_1$, $C_2$ and $C_3$, the blue region is a square region with side $L_x$ except the orange sites therein which form a smaller square with side $\ell_x$. The difference is obvious if we consider the shared boundary between the blue sites and the orange sites where in the second configuration from left is $4\ell_x$, in the third configuration from left is $3\ell_x$ and in the right one is $2\ell_x$.

It is worth to note that all of these configurations have corners which their contribution to logarithmic negativity for the case of $z=1$ has been previously studied in \cite{DeNobili:2016nmj}. Here we will not focus on the subleading corner contribution for generic $z$ and postpone this to future works.

\subsection{Area Law Behavior}
\begin{figure}
\begin{center}
\includegraphics[scale=0.28]{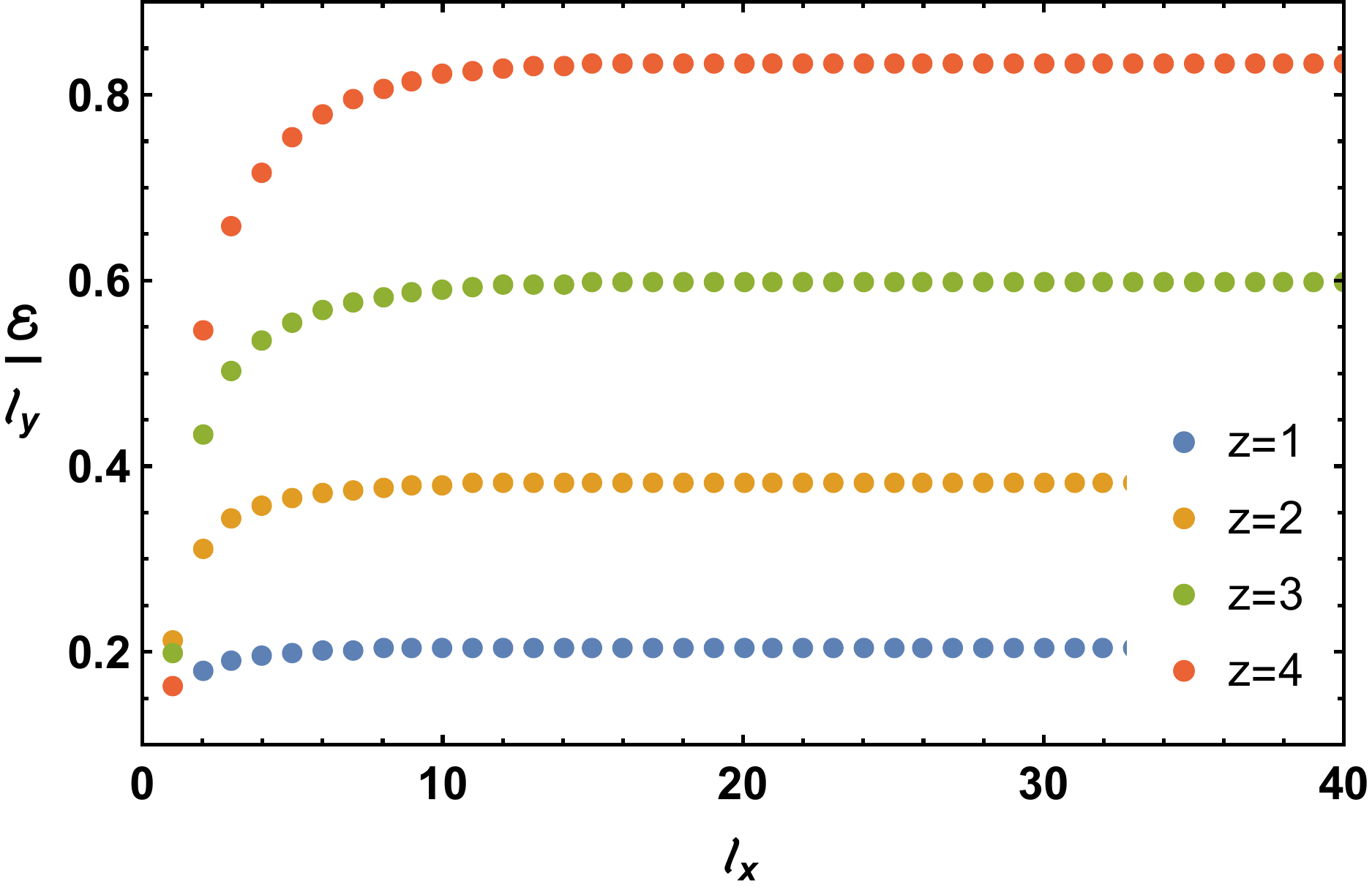}
\hspace{2mm}
\includegraphics[scale=0.27]{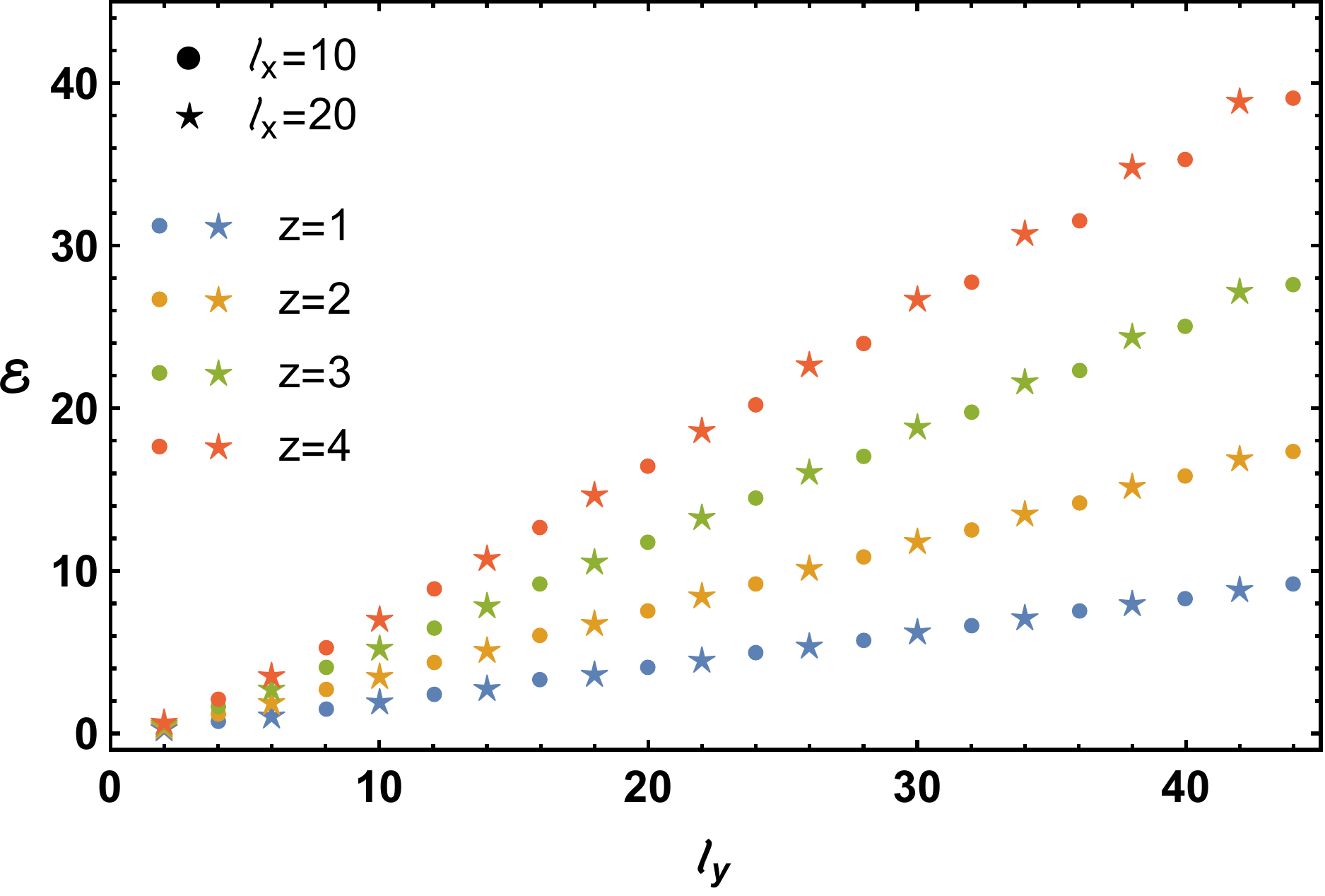}
\hspace{2mm}
\includegraphics[scale=0.27]{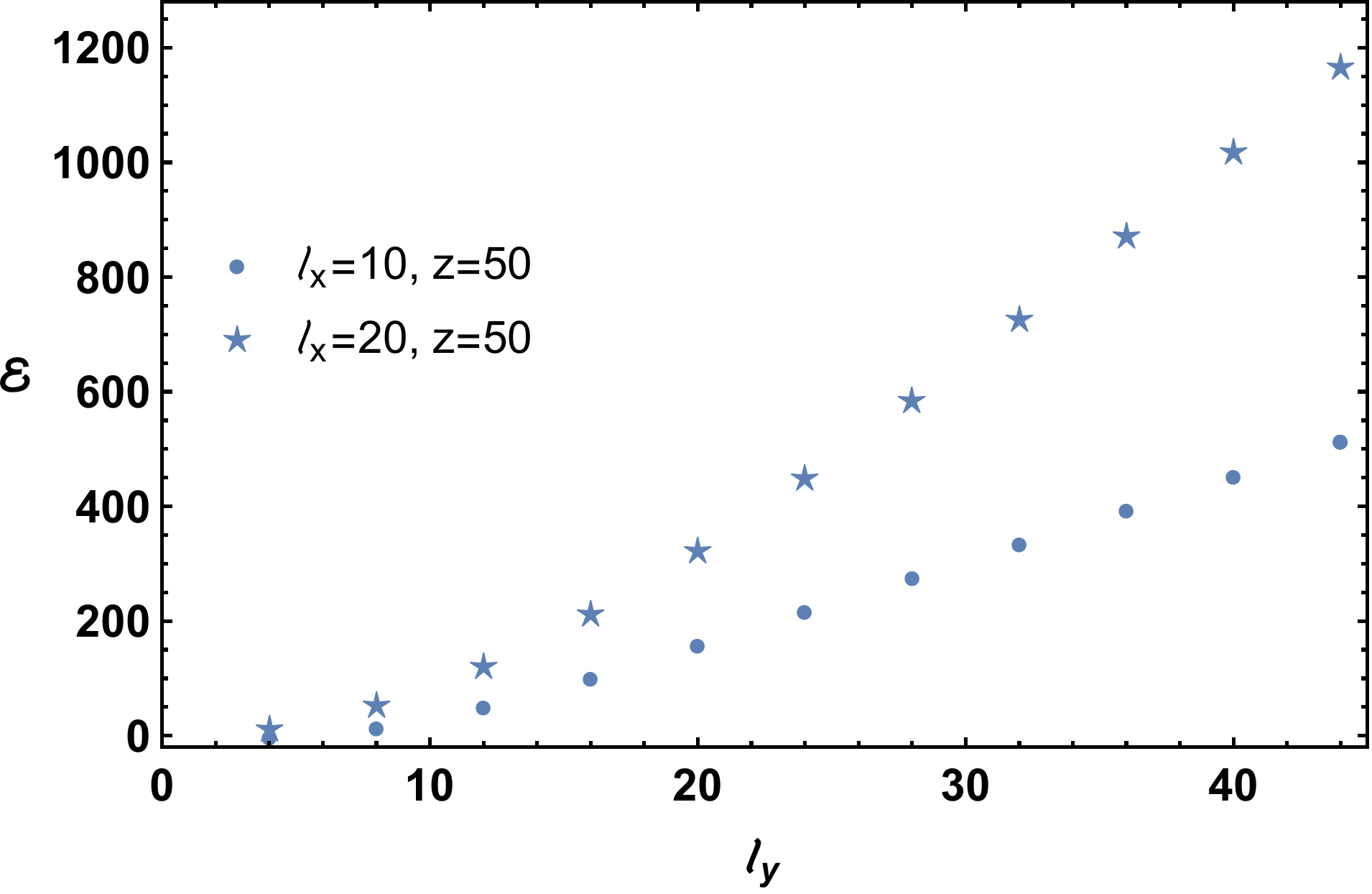}
\end{center}
\caption{Investigation of area law in logarithmic negativity for $z\ge 1$. In the left panel we have plotted the ratio $\mathcal{E}/\ell_y$ regarding to the left configuration in figure \ref{fig:configurations} with fixed value of $\ell_y=20$.
The middle panel shows logarithmic negativity for the same configuration again for $z=1,2,3,4$ for two different values of $\ell_x$ as a function $\ell_y$.  In our range of study for $z=1,2$, $\mathcal{E}$ grows with the same rate as $\ell_y$ is increased. For $z=3,4$ one can see that the curve corresponding to different values of $\ell_x$ start to dissever from each other as $\ell_y$ increases. This severance increases as $z$ increases. The right panel is showing the violation of shared area law in this model for $z=50$. In all panels we have set $N_x=N_y=300$ and $m=10^{-5}$.}
\label{fig:neg2p11}
\end{figure}

Here we focus on the behavior of logarithmic negativity for the left configuration in figure \ref{fig:configurations} where we construct a mixed state by tracing out the gray dots in that figure. We consider configuration $C_0$ where the region $A$ is divided into two symmetric subregions which are rectangles with width $\ell_x$ and height $\ell_y$ and we denote these rectangles by $A_1$ and $A_2$. The numerical data corresponding to this configuration for different values of dynamical exponent is plotted in the left panel of figure \ref{fig:neg2p11}. 

We have studied  $\mathcal{E}/\ell_y$ as a function of $\ell_x$ for different values of $\ell_y$. We have observed that for each value of $\ell_y$, while $\ell_x$ gets large enough, the value of $\mathcal{E}/\ell_y$ saturates to a finite value. This is showing that while we keep $\ell_y$ to be fixed, the localized degrees of freedom near the shared boundary between $A_1$ and $A_2$ which give the leading contribution to negativity do not increase by increasing $\ell_x$ and thus the value of negativity remains fixed. One can easily find that numerical results show that the value of $\ell_x$ which negativity reaches its maximum value increases with increasing the value of $\ell_y$ as expected. In the left panel of figure \ref{fig:neg2p11} we have plotted $\mathcal{E}/\ell_y$ versus $\ell_x$ for $\ell_y=20$. We have fixed the value of $\ell_y$ to prevent a messy plot.

The other important point which the numerical data is showing is that if we increase the value of the dynamical exponent (say the non-locality effects) in the model, for small enough dynamical exponents still the localized degrees of freedom near the boundary have the leading contribution to negativity but the number of these degrees of freedom increases with the value of $z$. We have shown the behavior for $z=1,2,3,4$ and in all cases for large enough $\ell_x$ the value of negativity saturates. This means that the expected area law still holds for this model but the thickness of the area contributing in the leading term of negativity increases with the dynamical exponent. A direct extrapolation of this shows that if the theory in consideration is on a finite lattice if the dynamical exponent increases this behavior should breakdown at some point. We will discuss about this in more details in what follows. 

In the middle panel of figure \ref{fig:neg2p11} we have plotted negativity as a function of $\ell_y$ for two different values of $\ell_x=10$ and $\ell_x=20$. The numerical data are showing that for this range of regions for $z=1$ and $z=2$ as expected from our above intuitive picture that the value of $\ell_x$ is not important and negativity for different values of $\ell_x$ coincide (i.e. the stars and the dots in the plot lay on the same line). On the other hand as the value of the dynamical exponent increases, the number of localized degrees of freedom contributing to the leading part of negativity increases. As a result of this, as $\ell_y$ increases, for different values of $\ell_x$ the number of points laying in the locality band near the shared boundary of $A_1$ and $A_2$ increases. Thus one can see that the negativity corresponding to different values of $\ell_x$ start to deviate from each other even for $z=3$ and $z=4$ and the deviation as expected is larger for larger values of $z$.

In the right panel of \ref{fig:neg2p11} we have plotted the numerical data corresponding to $z=50$ with the same values of $\ell_x$ and the same range for $\ell_y$. In this plot the two lines are totally separated from each other as a result of the band getting very thick which can not be called area law any more.

\begin{figure}
\begin{center}
\includegraphics[scale=0.27]{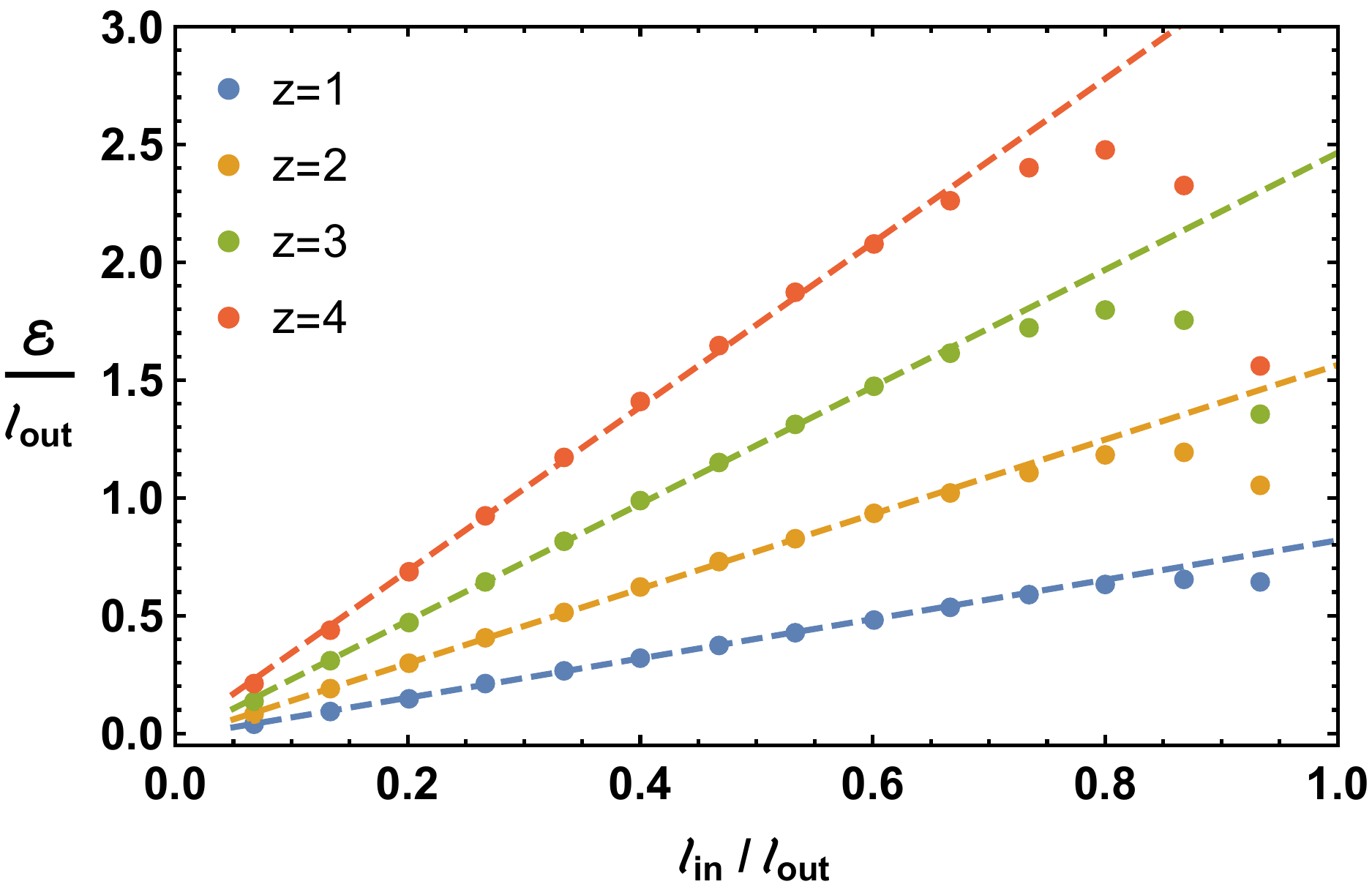}
\hspace{2mm}
\includegraphics[scale=0.26]{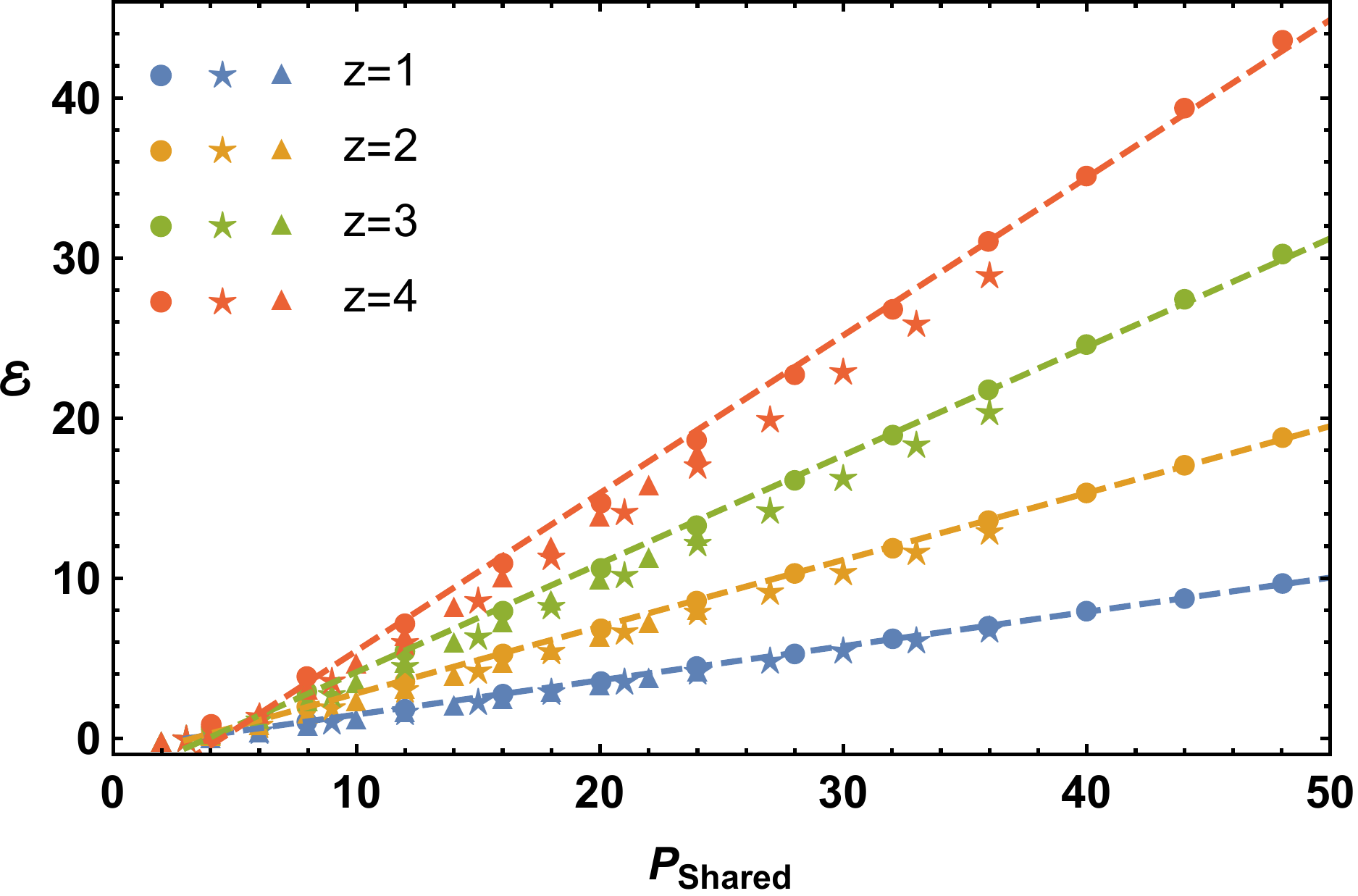}
\hspace{2mm}
\includegraphics[scale=0.27]{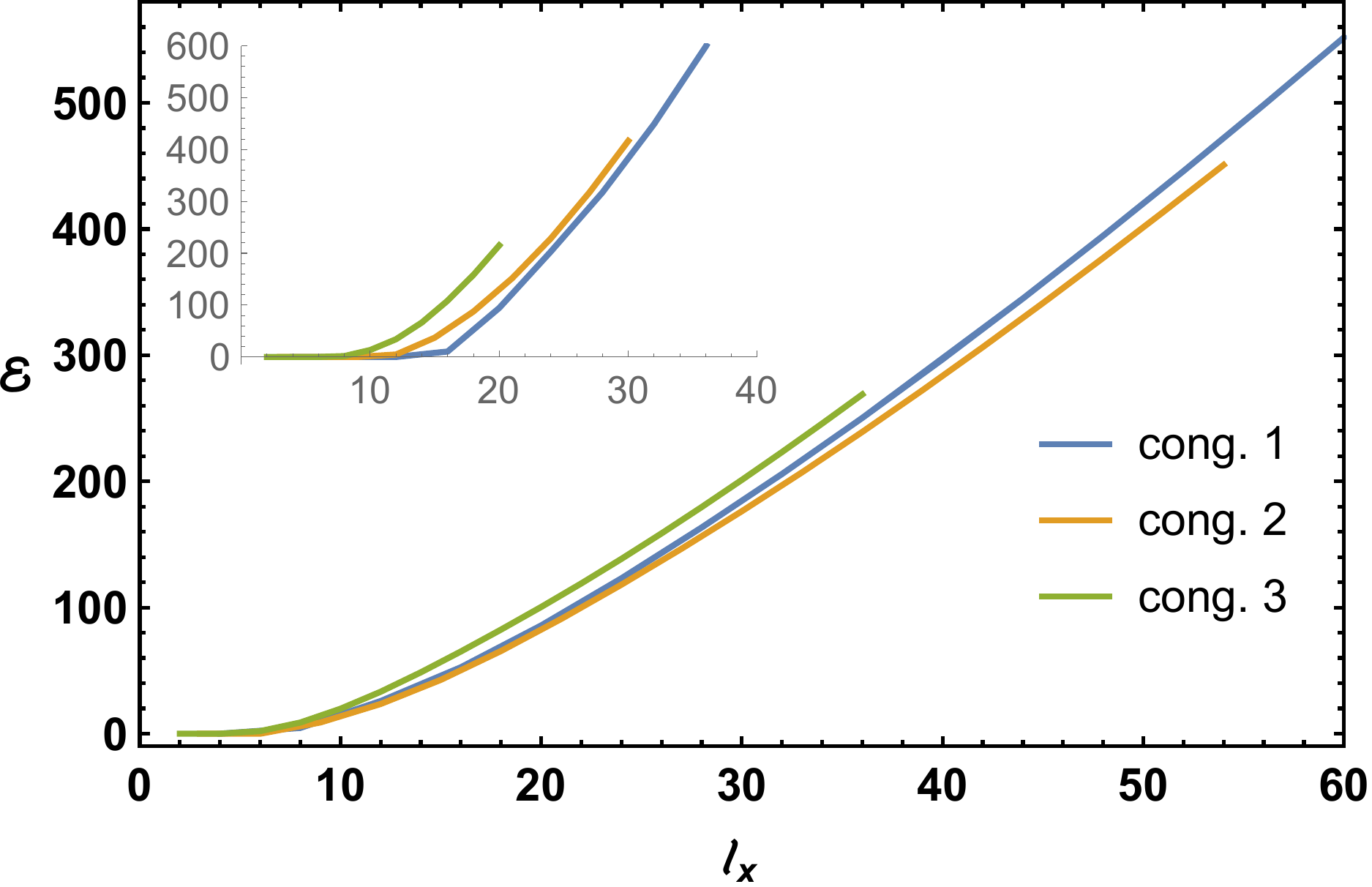}
\end{center}
\caption{The left panel shows logarithmic negativity for the second configuration (from left) in figure \ref{fig:configurations}. Here we have considered different values for the blue side and the orange side. The dashed line is also showing the fitting function for $\ell_{out}=30$ and $\ell_{in}/\ell_{out}=0.64$. The middle panel shows negativity for three right configurations shown in figure \ref{fig:configurations}. In these configurations we have set the side of the orange square as $1/3$ of the side of the blue square, i.e. the number sites inside $A_1$ is eight times of the number sites inside $A_2$. The dashed lines show the best fit for configuration 1, i.e. the second one from left in \ref{fig:configurations}. In the right panel we have shown the numerical data corresponding to the same configurations as the middle panel for $z=20$ in the major panel and for $z=50$ in the minor one.
In all panels we have set $N_x=N_y=300$ and $m=10^{-5}$.}
\label{fig:neg2p12}
\end{figure}
Now lets investigate the area law in configuration $C_1$ of figure \ref{fig:configurations}. Here we expect for small enough dynamical exponents, negativity should obey area law that is if we set $\ell_{out}$ (the side of the larger square with blue sites at the boundary) to be fixed, negativity should grow linearly as we increase $\ell_{in}$ (the side of the smaller square with orange sites at the boundary). In the left panel of figure \ref{fig:neg2p12} we have plotted this for several (small enough) values of the dynamical exponent and $\ell_{out}=30$. We should note that since $0<\ell_{in}<\ell_{out}$, as $\ell_{in}\to\ell_{out}$ negativity should vanish since in this case the subregion $A_1$ vanishes. Thus as we expect, for each value of small dynamical exponent, after $\ell_{in}$ gets greater than a critical value, negativity starts to descend and finally vanishes where $\ell_{in}=\ell_{out}$. Before this critical value negativity obeys the expected area law.


As another evidence for the area law we compare the behavior of negativity for three different configurations in the right of figure \ref{fig:configurations}. For each of these configurations the number of sites in subregion $A_1$ and $A_2$ are kept fixed but the place of $A_2$ varies such that the shared boundary is different. The numerical data plotted in the middle and right panel of figure \ref{fig:neg2p12} correspond to configurations that the side of the largest side of $A_1$ is always 3 times the side of $A_2$. Again in the middle panel we have plotted data corresponding to $z=1,2,3,4$. The blue part corresponds to $z=1$. In this case one can easily see that the behavior of negativity is the same for different configurations. They all grow linearly with almost the same slope which is a strong evidence for area law, in terms of the shared boundary between $A_1$ and $A_2$. The orange data corresponding to $z=2$ show almost the same behavior although some deviation from a single line is observed even in this case. The deviation gets larger as $z$ increases. This can be obviously seen in the green and red data corresponding to $z=3,4$. 

Beside from violation of area law, our numerical data also shows an `initial sleep' regime similar to what we reported in the $(1+1)$-dimensional case. Again as the dynamical exponent increases, there is a regime which the logarithmic negativity is zero for subregions smaller than a certain value. The right panel of figure \ref{fig:neg2p12} we have plotted our numerical data corresponding to $z=20$ and $z=50$ and all other parameters fixed. As one can see by comparing the main panel corresponding to $z=20$ with the internal panel corresponding to $z=50$, the length of this regime increases with the dynamical exponent. We believe that this behavior is a lattice effect similar to the $(1+1)$-dimensions.

\subsection*{Finite Temperature}

\begin{figure}
\begin{center}
\includegraphics[scale=0.39]{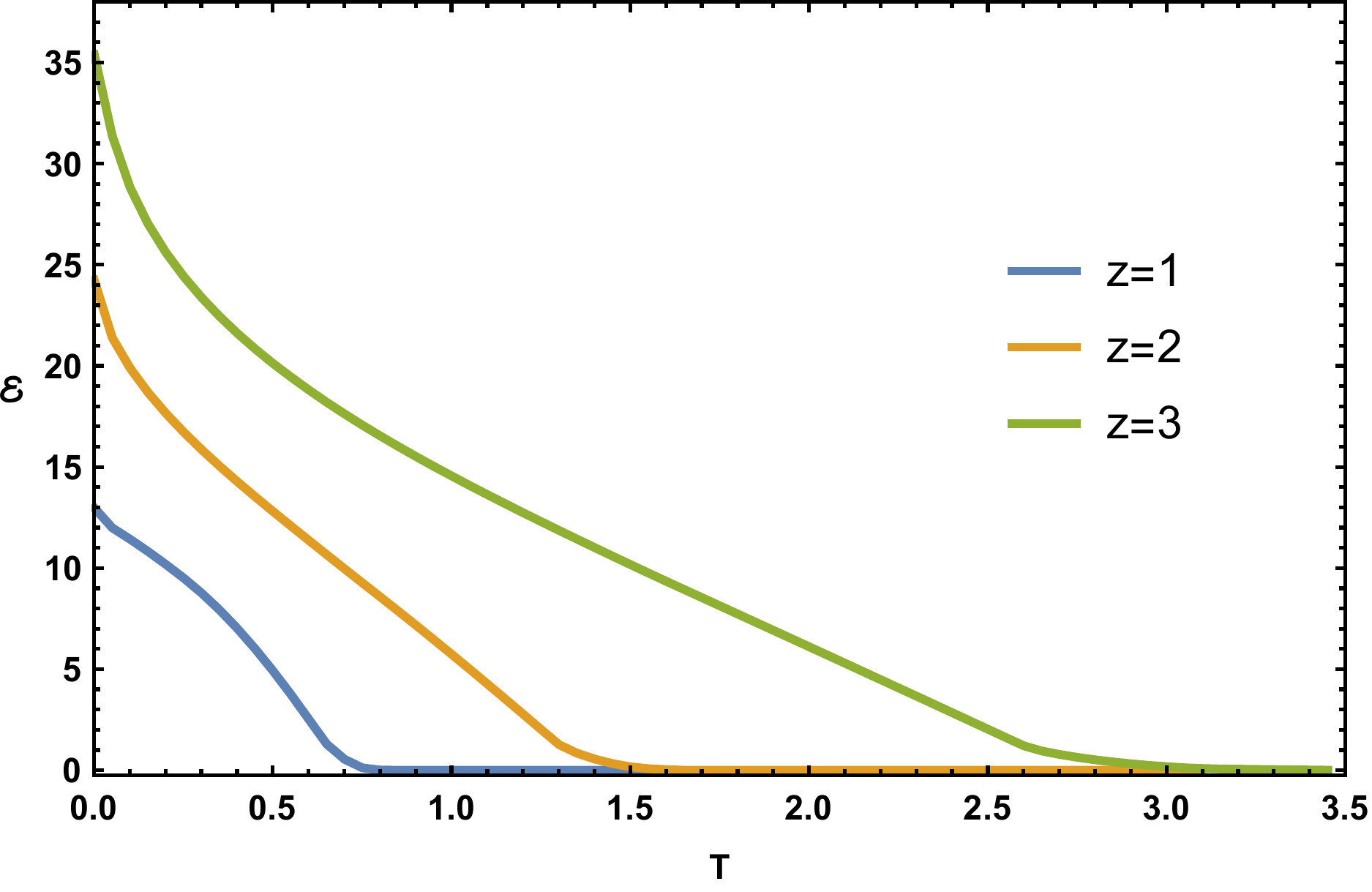}
\hspace{5mm}
\includegraphics[scale=0.4]{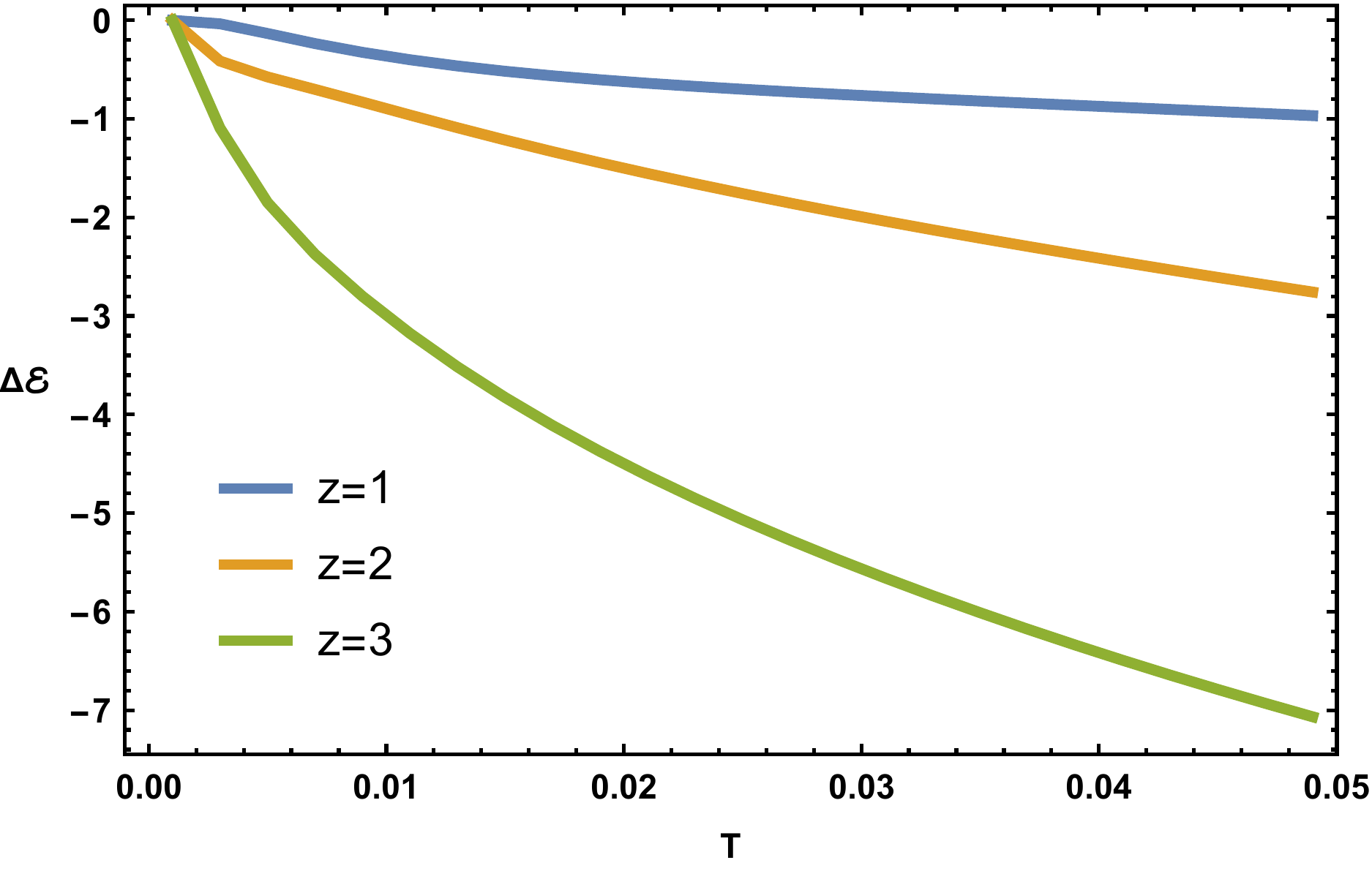}
\end{center}
\caption{Logarithmic negativity for finite temperature states in $(2+1)$-dimensions. Here we have considered a square with side $\ell_x=\ell_y=15$ as $A_1$ and the rest of a square lattice with $N_x=N_y=40$ as $A_2$. The left panel shows the behavior of logarithmic negativity between $T=0$ and $T_{sd}$ where the sudden death phenomena happens. In the right panel we have plotted the subtracted logarithmic negativity corresponding to very small temperatures far from $T_{sd}$.}
\label{fig:negfT2p1}
\end{figure}
The generic behavior of logarithmic negativity in finite temperature states in $(2+1)$-dimensions is very similar to what we have described in $(1+1)$-dimensional case. The decreasing behavior as a function of temperature and the sudden death phenomena happens here in a very similar manner which we are not going to repeat here. In the left panel of figure \ref{fig:negfT2p1} we have plotted the corresponding behavior in $(2+1)$-dimensions and in the right panel of the same figure we have plotted the subtracted logarithmic negativity in the low-temperature regime.

\subsection*{$z$-dependence of negativity}
Similar to what we did in $(1+1)$-dimensions here we would like to study how does logarithmic negativity depend on the dynamical exponent. As we have seen in $(1+1)$-dimensions and also in other investigations in this section for $(2+1)$-dimensions, negativity increases with $z$. To do so we have plotted our numerical data regarding to the vacuum state in the left panel of figure \ref{fig:negZdep1p1}. The behavior is very similar to what we have reported in the previous section for $(1+1)$-dimensions and after a short range of slow increase logarithmic negativity increases linearly with $z$. This is also similar to the $z$-dependence of entanglement entropy in  $(2+1)$-dimensions which was previously studied in \cite{MohammadiMozaffar:2017nri}. 

In the left panel we have plotted the data corresponding to the $z$-dependence of logarithmic negativity in states at finite temperature. In this case we have plotted the data regarding to the regime comparable with field theory, i.e., we have considered $T\ll 1$. The maximum value of the temperature is set to be $T=10^{-2}$. In this regime we have found that as the temperature increases, the non-linear regime extends to be valid for a larger range of $z$ but again it enters the linear regime. This is very similar to what we have reported in the case of $(1+1)$-dimensions both for extended and for the $p$-alternating subregions.
\section{Conclusions and Discussions}\label{sec:conclusions}
We have mainly studied the logarithmic negativity for a specific harmonic model with Lifshitz scaling symmetry as an extension of our previous work\cite{MohammadiMozaffar:2017nri}. The main practical feature of this quantity which is believed to be a suitable entanglement measure for mixed states is that it is a computable measure. We have studied different aspects of logarithmic negativity in $(1+1)$ and $(2+1)$-dimensions in various setups and with different boundary conditions numerically. Also in order to gain more insights into certain properties of this quantity we have considered an specific lattice configuration the so-called $p$-alternating sublattice introduced in \cite{He:2016ohr} which is analytically tractable, and thus offers more concrete results. 

\begin{figure}
\begin{center}
\includegraphics[scale=0.4]{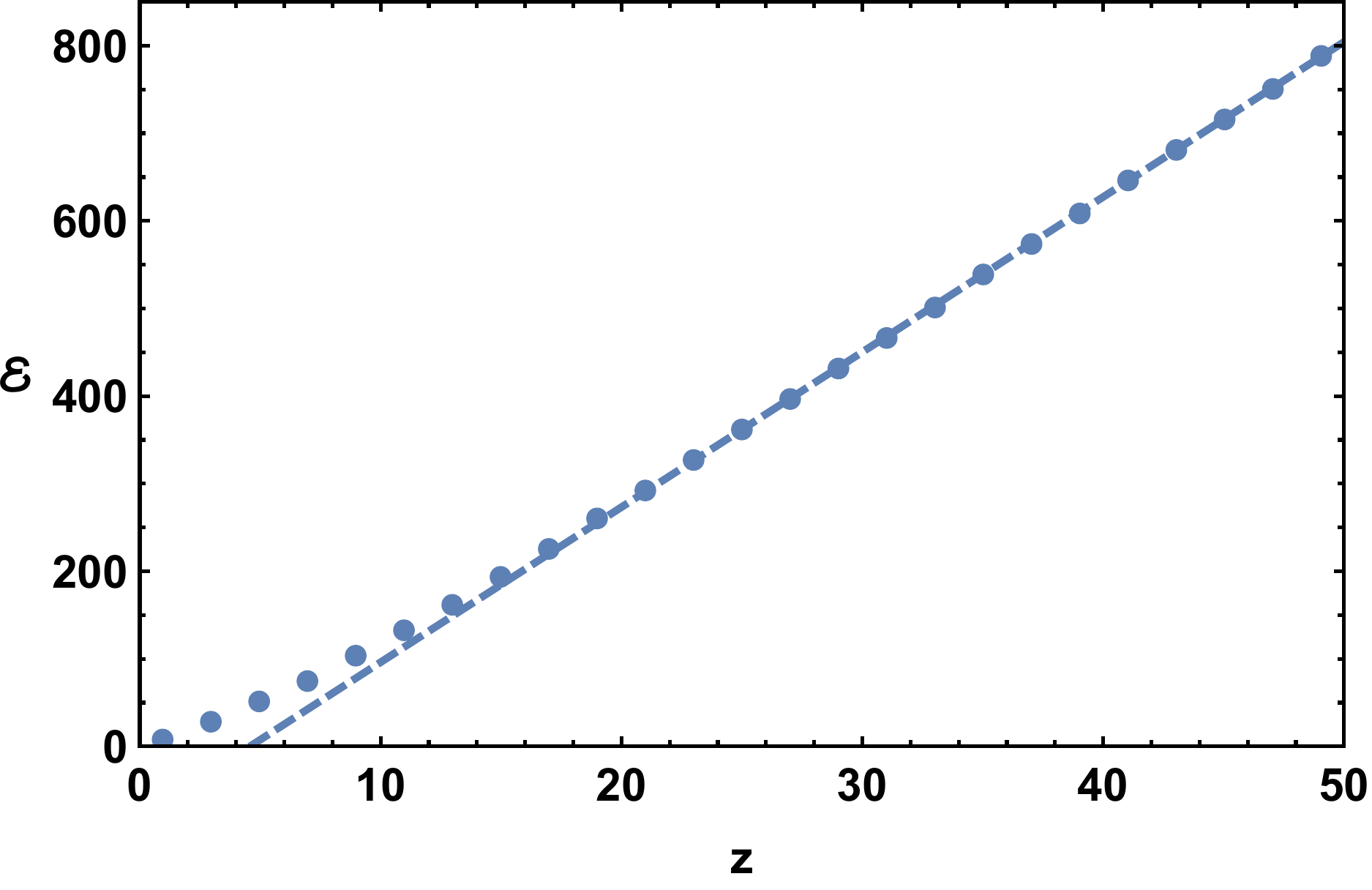}
\hspace{5mm}
\includegraphics[scale=0.4]{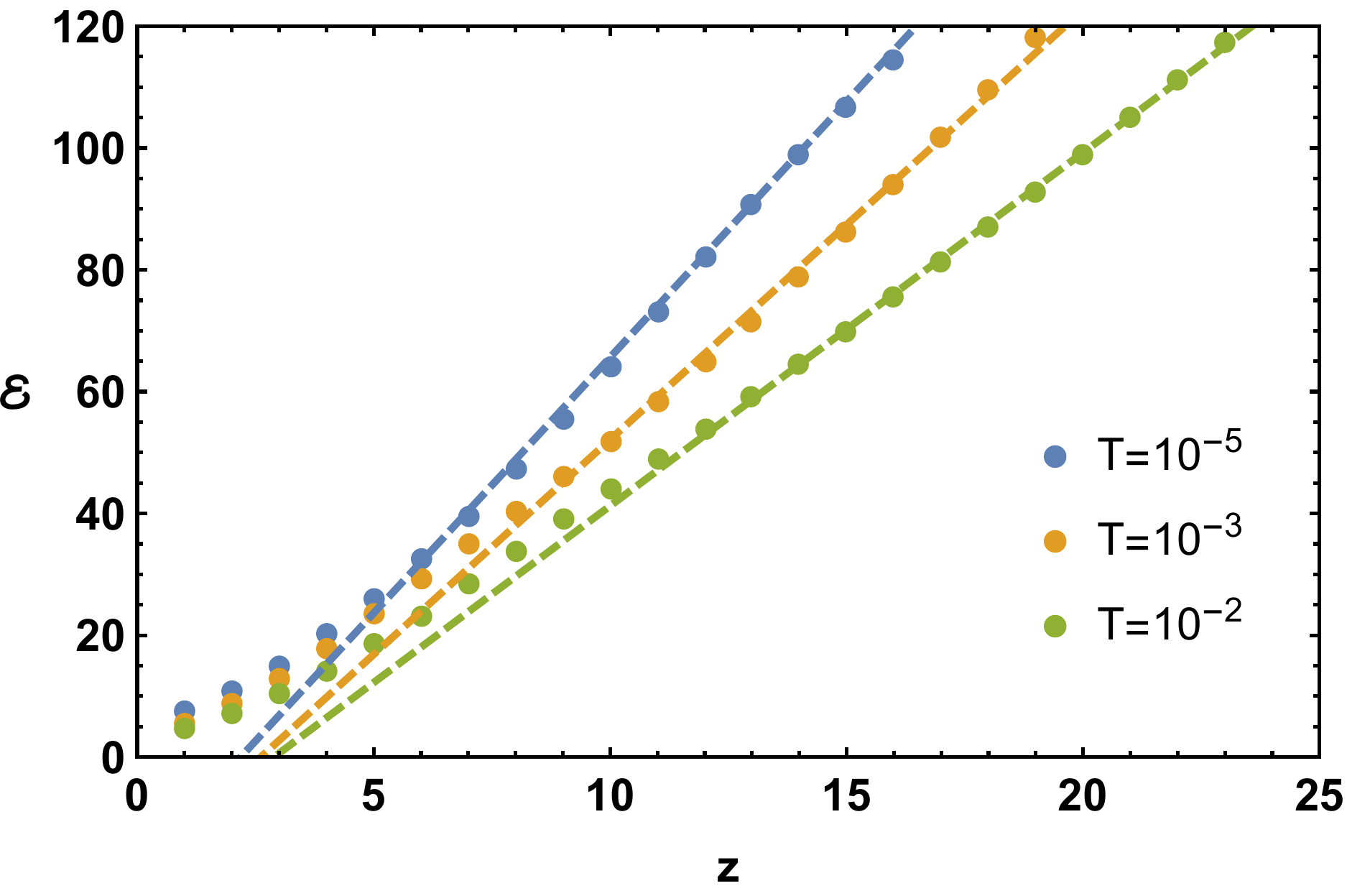}
\end{center}
\caption{Dynamical exponent dependence of logarithmic negativity at vacuum and finite temperature states. The left panel is showing data corresponding to the vacuum state. The data corresponds to configuration $C_3$ with $L_x=12$ and $\ell_x=6$. The right panel corresponds to finite temperature states with different temperatures. In both plots we have set $m=10^{-5}$}
\label{fig:negZdep2p1}
\end{figure}

In the following we would like to summarize our main results.

\begin{itemize}
\item In the case of $(1+1)$-dimensional systems with Lifshitz scaling symmetry for configuration $C_2$, we have shown that for small enough dynamical exponents, i.e., $0<z<2$, the logarithmic negativity is a logarithmic function of the length of the entangling region. In the regime which we expect approximation of the continuum, using the numerical data and employing a suitable fit function, we have found a universal coefficient which we denoted by $c_{eff}$. In the conformal case, i.e., $z=1$, this universal coefficient excellently coincides with the central charge of 2-dimensional bosonic CFT. Our results also show that for $0<z<2$ the effect of the mass parameter is packed in the non-universal part of logarithmic negativity.

\item In the case of thermal mixed states in $(1+1)$-dimensions, we have found that the logarithmic negativity is a
decreasing function of temperature for all values of the dynamical exponent. Such a behavior is in agreement with our expectation since at high temperatures, a quantum system crossovers to a classical one. As the temperature is increased, logarithmic negativity is well-known to vanish for $T > T_{sd}$ due to lattice construction. We have shown that for larger values of $z$, $T_{sd}$ increases. We believe that this peculiar behavior is a numerical artifact and it is not relevant to the continuum limit. In order to avoid this lattice effect we have focused on small temperature regime and we have shown that the magnitude of $\mathcal{E}(T)-\mathcal{E}(0)$ increases while the dynamical exponent is increased.

\item Focusing on the $z$-dependence of logarithmic negativity, our numerical data shows that as the dynamical exponent increases, the value of logarithmic negativity also increases. Similar to the case of entanglement entropy \cite{MohammadiMozaffar:2017nri} we can understand this intuitively in terms of increasing
number of correlated sites on the lattice in the Lifshitz harmonic lattice. Our results show that in both $(1+1)$ and $(2+1)$-dimensions for small values of the dynamical exponent logarithmic negativity grows faster than linear with $z$ and for large enough dynamical exponent  it grows linearly with $z$. This behavior seems to be independent of the properties of the entangling region and also independent of space-time dimensions.

\item Another $(1+1)$-dimensional setup that we have considered is an specific configuration known as $p$-alternating
sublattice on a periodic lattice. The advantage of studying this configuration is that the correlator matrices in such systems become circulant matrices which their eigenvalues and eigenvectors are known analytically. In this case we have shown that for a specific configuration, similar to entanglement entropy, the partial transpose matrix remains circulant and thus we can perform calculation of logarithmic negativity analytically.
For these configurations we have shown that the logarithmic negativity is a linear function of the dynamical exponent which agrees with our numerical results corresponding to extended configurations. Also we have found the thermal corrections to this quantity analytically.

\item In the case of $(2+1)$-dimensions we have investigated the existence of logarithmic negativity area law which in other words says that logarithmic negativity scales with the boundary shared by $A_1$ and $A_2$. Our numerical investigation shows that in the presence of the dynamical exponent, for small values of $z$ in comparison with the characteristic length of the region in consideration in units of the lattice spacing, the shared area law remains valid but for larger values of $z$ our data shows deviation from this area law and the deviation becomes larger as we increase the value of $z$.

\item We have observed a new kind of behavior for entanglement negativity in both  $(1+1)$ and $(2+1)$-dimensions. For small subsystems entanglement negativity vanishes and as the length of the region in consideration is increasing, entanglement negativity starts to get non-vanishing values for systems larger than a characteristic length which grows with the dynamical exponent. We have shown that this `initial sleep' regime is a lattice effect and it disappears in the continuum limit. 

\end{itemize}

There are several directions which one can follow to further investigate different aspects of quantum entanglement for  models with Lifshitz scaling symmetry. We leave further investigations of these
aspects, including the time evolution of entanglement entropy after a global quantum quench, to future works.

\section*{Acknowledgments}
We would like to thank Mohsen Alishahiha and Mohammad Ali Rajabpour for their useful comments on an early draft of this work. We are also very grateful to Andrea Coser for correspondence, careful reading of the manuscript and his valuable comments.


\end{document}